\documentclass[12pt]{article}
\usepackage[top=1in,left=1in,bottom=1in,right=1in, paperheight=11in, paperwidth=8.5in]{geometry}

\usepackage{adjustbox}
\usepackage{sectsty}
\usepackage[hyphens]{url}
\usepackage[breaklinks]{hyperref}
\usepackage{amsmath, amssymb, amsthm, amsbsy, amsfonts, cases}
\usepackage{graphicx, comment}
\usepackage[noend]{algpseudocode}
\sectionfont{\fontsize{13}{13}\selectfont}
\subsectionfont{\fontsize{12}{12}\selectfont}
\usepackage{bigints}
\usepackage[mathscr]{euscript}
\usepackage[makeroom]{cancel}
\usepackage{graphicx}
\usepackage[table]{xcolor}
\usepackage{subcaption}
\usepackage{enumitem}
\usepackage{verbatim}
\usepackage[font={small,it,singlespacing}]{caption}
\usepackage{lscape}
\usepackage{caption}
\usepackage{hyperref}
\usepackage[sort]{natbib}
\usepackage{parskip}
\usepackage{array}
\usepackage{multirow}
\usepackage{graphicx}
\usepackage{wrapfig}
\usepackage{lscape}
\usepackage{rotating}
\usepackage{epstopdf}
\usepackage[compact]{titlesec}
\usepackage{makecell}
\usepackage{authblk}
\usepackage[toc,page]{appendix}

\usepackage{color, colortbl}
\definecolor{Gray}{gray}{0.9}

\usepackage{algorithm}% http://ctan.org/pkg/algorithm
\usepackage{algpseudocode}% http://ctan.org/pkg/algorithmicx

\usepackage{boldline}
\hypersetup{colorlinks,citecolor=blue}
\graphicspath{{./images//}}

\newcommand{\M}{\mathcal{M}}

\newcommand{\x}{\mathbf{x}}

\newcolumntype{L}[1]{>{\raggedright\let\newline\\\arraybackslash\hspace{0pt}}m{#1}}
\newcolumntype{C}[1]{>{\centering\let\newline\\\arraybackslash\hspace{0pt}}m{#1}}
\newcolumntype{R}[1]{>{\raggedleft\let\newline\\\arraybackslash\hspace{0pt}}m{#1}}

\theoremstyle{plain}
\newtheorem{thm}{Theorem}
\newtheoremstyle{exampstyle}
  {\topsep} % Space above
  {\topsep} % Space below
  {} % Body font
  {} % Indent amount
  {\bfseries} % Theorem head font
  {} % Punctuation after theorem head
  {.5em} % Space after theorem head
  {} % Theorem head spec (can be left empty, meaning `normal')
 
\newtheorem{defn}{Definition}
\newtheoremstyle{exampstyle}
  {\topsep} % Space above
  {\topsep} % Space below
  {} % Body font
  {} % Indent amount
  {\bfseries} % Theorem head font
  {.} % Punctuation after theorem head
  {.5em} % Space after theorem head
  {} % Theorem head spec (can be left empty, meaning `normal')

\theoremstyle{exampstyle}
\theoremstyle{exampstyle}
\theoremstyle{exampstyle}
\theoremstyle{exampstyle}
\theoremstyle{exampstyle}\newtheorem{rem}{Remark}

\providecommand{\keywords}[1]
{
  \small	
  \textbf{\textit{Keywords---}} #1
}

\title{\normalsize{\textbf{Comparative Study of Differentially Private Synthetic Data Algorithms from the NIST PSCR Differential Privacy Synthetic Data Challenge}}}
\author[1]{\small{Claire McKay Bowen}}
\author[2]{\small{Joshua Snoke}}
\affil[1]{Urban Institute\\ cbowen@urban.org}
\affil[2]{RAND Corporation\\ jsnoke@rand.org}

\date{}

\begin{document}
\maketitle

\vspace{-0.5cm}

\begin{abstract}
    \noindent Differentially private synthetic data generation offers a recent solution to release analytically useful data while preserving the privacy of individuals in the data. In order to utilize these algorithms for public policy decisions, policymakers need an accurate understanding of these algorithms' comparative performance. Correspondingly, data practitioners require standard metrics for evaluating the analytic qualities of the synthetic data. In this paper, we present an in-depth evaluation of several differentially private synthetic data algorithms using actual differentially private synthetic data sets created by contestants in the 2018-2019 National Institute of Standards and Technology Public Safety Communications Research (NIST PSCR) Division's ``Differential Privacy Synthetic Data Challenge.'' We offer analyses of these algorithms based on both the accuracy of the data they created and their usability by potential data providers. We frame the methods used in the NIST PSCR data challenge within the broader differentially private synthetic data literature. We implement additional utility metrics, including two of our own, on the differentially private synthetic data and compare mechanism utility on three categories. Our comparative assessment of the differentially private data synthesis methods and the quality metrics shows the relative usefulness, the general strengths and weaknesses, and offers preferred choices of algorithms and metrics. Finally we describe the implications of our evaluation for policymakers seeking to implement differentially private synthetic data algorithms on future data products. 
\end{abstract}

    % We also investigate two of our own original, previously published differentially private synthetic data algorithms to evaluate their comparative performance, framing both the NIST data challenge methods and our own algorithms within the broader differentially private synthetic data literature. 
    
\section{Introduction}\label{S:one}

\subsection{Background on Differentially Private Synthetic Data}
The collection and dissemination of data can greatly benefit society by enabling a range of impactful research projects, such as the Personal Genome Project Canada database which determines Genomic variants in participants for several health problems \citep{reuter2018personal}, the United Kingdom Medical Education Database ``to improve standards, facility workforce planning and support the regulation of medical education and training'' \citep{dowell2018uk}, and the Robert Wood Johnson Foundation 500 Cities Project that provided a large United States data set that ``contain[ed] estimates for 27 indicators of adult chronic disease, unhealthy behaviors, and preventative care available'' as a ``groundbreaking resource for establishing baseline conditions, advocating for investments in health, and targeting program resources where they are needed most'' \citep{500cities}. However, sharing data based on human subjects with potentially sensitive information often raises valid concerns over the privacy risks inherit in sharing data, and recent misuses of data access for seemingly research purposes, such as the Facebook - Cambridge Analytica Scandal, have heightened data privacy concerns over how private companies and government organizations gather and disseminate information \citep{martin2017data, tsay2018social, gonzalez2019global}.

Statistical disclosure control (SDC), or limitation (SDL), exists as a field of study that aims to develop methods for releasing high-quality data products while preserving the privacy and confidentiality of sensitive data. These techniques have existed within statistics, the social sciences, and government agencies since the mid-twentieth century, and they have sought to balance risk against the benefit to society, also known as the utility of the data (we will use quality and utility interchangeably in this paper). SDC methods traditionally require strong assumptions concerning the knowledge and identification strategies of the attacker, and risk is estimated by simulating these attackers based on those assumptions \citep{reiter2005using, hundepool2012statistical, manrique2012estimating}.
%This issue is a common critique for SDC methods that rely on specific data values and assumptions on the data adversary's behavior and background knowledge. 

While this field has existed for some time, over the past two decades the data landscape has dramatically changed. With the growth of the internet and tech companies, data collection and data sharing have vastly increased. Data adversaries (also referred to as intruders or attackers) can more easily reconstruct data sets and identify individuals from supposedly anonymized data using advances in modern information infrastructure and computational power. Examples of re-identified anonymized data include the Netflix Prize data set \citep{narayanan2008robust}, the Washington State health records \citep{sweeney2013matching}, credit card metadata \citep{de2015unique}, cell phone spatial data \citep{hardesty2013hard, de2013unique, kondor2018towards}, and the United States public use microdata files \citep{rocher2019estimating}. Due to the increased availability of external data files and methods for reconstructing information from data, data privacy practitioners have dwindling confidence that they can accurately assess the risk of releasing data based on simulating all plausible adversaries.

Beginning with \cite{dwork2006calibrating}, researchers have begun developing a new concept, known as differential privacy (DP) or sometimes more generally as formal privacy, to combat this heightened risk of privacy loss. Privacy researchers consider DP the first provable definition for quantifying privacy loss when releasing information from a confidential data set. In contrast to prior SDC methods, DP does not require a simulated attacker or the same strong assumptions concerning how much information an intruder may have or what kind of disclosure is likely to occur. This does not imply that DP protects from all attacks, but for a defined type of privacy loss it offers provable amounts of protection. At a high level, DP links the potential for privacy loss to how much the answer of a query (such as a statistic) is changed given the absence or presence of the most extreme possible person in the population of the data. DP requires that the level of protection is set proportional to this maximum potential change, thereby providing formal privacy protections scaled to the worst-case scenario. For further details, \cite{dwork2014algorithmic} provides a rigorous mathematical review of DP while \cite{nissim2017differential} and \cite{snoke2019differential} describe DP for a non-technical, general audience. Since its conception, DP has created an entire new field of research with applications in Bayesian learning \citep{wang2015privacy}, data mining \citep{mohammed2011differentially}, data streaming \citep{dwork2010pan}, dimension reduction \citep{chaudhuri2012near}, eye tracking \citep{liu2019differential}, genetic associate tests \citep{yu2014scalable}, inferential statistical analyses \citep{karwa2016inference, wasserman2010statistical}, power grid obfuscation \citep{fioretto2019differential}, and recommender systems \citep{friedman2016differential} to list a few.

Synthetic data generation is another innovation in SDC that has become a leading practical approach for releasing publicly available data that can be used for exploratory purposes and numerous different analyses \citep{rubin1993statistical, little1993statistical, raghunathan2003multiple, reiter2005using, drechsler2011synthetic, raab2016practical}. While this approach has been shown to offer improvements in preserving the utility of the data compared against other SDC methods, as originally proposed it lacks a formal privacy quantification.

Within the DP literature, researchers have more recently considered the combination of DP and data synthesis as a solution to releasing analytically useful data while preserving the privacy of individuals in the data. Applications include binary data \citep{charest2011can, mcclure2012differential}, categorical data \citep{abowd2008protective, hay2010boosting}, continuous data \citep{wasserman2010statistical,BowLiu2020,snoke2018pmse}, network data \citep{karwa2016inference, karwa2017sharing}, and Poisson distributed data \citep{quick2019generating}. In order to utilize these algorithms for public policy decisions, policymakers need an accurate understanding of these algorithms' comparative performance. However, there are very few studies comparing multiple differentially private data synthesis methods and, to the best of our knowledge, no studies applying the comparisons on real-world data. Correspondingly, data practitioners wishing to produce differentially private synthetic data are unlikely to know what algorithms fit their application or find information concerning the relative strengths and weaknesses of different approaches.
%require standard metrics for evaluating the analytic qualities of the synthetic data.
%Guidance on what utility or quality metrics to implement is also lacking in the DP literature.

\subsection{Contributions of This Paper}
In this paper, we provide an in-depth assessment of various differentially private data synthesis methods applied to multiple real, non-trivial data sets and evaluated on a variety of utility metrics. The underlying algorithms and data for this study comes from the 2018 National Institute of Standards and Technology Public Safety Communication Research (NIST PSCR) Division's ``Differential Privacy Synthetic Data Challenge'' \citep{vendetti20182018}. Due to the competitive nature of the challenge, final scores had to be aggregated such that more academic evaluations of the algorithms' performances were not possible. 

In our assessment, we evaluate the differentially private data synthesis mechanisms based on their performance in the challenge and on a wider range of utility metrics. We provide descriptions of each algorithm and consider their ease of implementation, such as the availability of open-source code, computational feasibility, and the amount of public data pre-processing required. We also note algorithms' current standing as published methods in the data privacy and confidentiality literature.

We expand on the scoring metrics devised for the challenge and evaluate the synthetic data sets on a variety of other standard metrics in the data privacy utility literature. For readers unfamiliar with different ways to evaluate accuracy, this paper offers a concise and well-organized set of metrics that form a broad utility assessment. We organize the utility metrics used to assess the synthetic data in one of three groups: (1) marginal distribution metrics, (2) joint distribution metrics, and (3) correlation metrics. Using the actual synthetic data sets generated by contestants in the challenge, which were made available to us by NIST PSCR, we implemented multiple metrics in each of these categories. We also assigned each of the three NIST PSCR Data Challenge scoring measures to one of these categories. 

The categories provide our evaluation with multiple lenses on the relative usefulness of each algorithm, ranging from specific measures of model accuracy to general measures of distributional similarity. We provide recommendations for best candidate methods, the first such recommendations based on a large-scale real data application, for future use based on their strengths and weaknesses. We provide policymakers and practitioners seeking to implement differentially private synthetic data algorithms both an assessment of the algorithms used in the challenge and a framework for evaluating future differentially private synthetic data techniques. 

We organize the remainder of the paper as follows. Section \ref{sec:DP} reviews the definitions and concepts of differential privacy and common differentially private mechanisms. Section \ref{sec:DPDS} summarizes the differentially private data synthesis methods ranked in the NIST PSCR Data Challenge, and Section \ref{sec:DP-Metric} describes the quality metrics we implemented on the NIST PSCR Data Challenge data sets. Section \ref{sec:study} evaluates and compares all the quality metric results. Concluding remarks and suggestions for future work are given in Section \ref{sec:discussion}.

\section{Differential Privacy}\label{sec:DP}
Differential privacy (DP), under a specific definition, offers a provable and quantifiable amount of privacy protection, colloquially referred to as the privacy-loss budget. Satisfying DP is a statement about the algorithm (or mechanism), not a statement about the data. Rather than stating that the output data meets privacy requirements, DP requires that the \textit{algorithm} which produces the output provably meets the definitions. Accordingly, algorithms which satisfy the definitions are referred to as differentially private algorithms.

In this section, we reproduce the pertinent definitions and theorems of DP with the following notation: $X\in\mathbb{R}$ is the original data set  with dimension ${n\times q}$ and $X^*$ is the private version of $X$ with dimension ${n^*\times q}$. We also define a statistical query as a function $u:\mathbb{R}^{n\times q}\rightarrow\mathbb{R}^k$, where the function maps the possible data sets of $X$ to $k$ real numbers.

\subsection{Definitions and Theorems}\label{subsec:def}
\begin{defn}\label{def:dp} \textbf{Differential Privacy} \citep{dwork2006calibrating}:
A sanitization algorithm, $\M$, gives $\epsilon$-DP if for all subsets $S\subseteq Range(\M)$ and for all $X,X'$ such that $d(X,X')=1$, 
    \begin{equation}\label{eqn:dp}
        \frac{\Pr(\M( X) \in S)}{ \Pr(\M( X')\in S)}\le \exp(\epsilon)
    \end{equation}
\end{defn}
\noindent where $\epsilon>0$ is the privacy-loss budget and $d(X,X')=1$ represents the possible ways that $X'$ differs from $X$ by one record. We define this difference as a presence or absence of a record, but note that some definitions of DP has this difference as a change, where $X$ and $X'$ have the same dimensions.

One common concern about algorithms that satisfy $\epsilon$-DP is they tend to inject a large amount of noise into statistical query results. Several relaxations of $\epsilon$-DP have been developed such as approximate DP \citep{dwork2006our}, probabilistic DP \citep{machanavajjhala2008privacy}, and concentrated DP \citep{dwork2016concentrated}. These are called relaxations because, while still formal, they offer slightly weaker privacy guarantees. In return, they typically lessen the amount of noise required. We will cover approximate DP, also known as $(\epsilon, \delta)$-DP, since the NIST PSCR Data Challenge allowed the submissions to satisfy $(\epsilon, \delta)$-DP rather than strict $\epsilon$-DP.

\begin{defn}\label{def:adp} \textbf{$(\epsilon, \delta)$-Differential Privacy} \citep{dwork2006our}:
A sanitization algorithm $\M$ gives $(\epsilon, \delta)$-DP if for all $X, X'$ that are $d(X,X')=1$,
    \begin{equation}\label{eqn:adp}
        \Pr(\M( X) \in S)\le \exp(\epsilon) \Pr(\M( X')\in S) + \delta
    \end{equation}
    where $\delta\in [0,1]$. $\epsilon$-DP is a special case of $(\epsilon, \delta)$-DP when $\delta=0$.
\end{defn}

The parameter $\delta$ adds a small probability that the bound given in Definition \ref{def:dp} does not hold, which can be useful when dealing with extreme yet very unlikely cases. 

%In general, if the data practitioner (or curator) ``spends'' more of the privacy-loss budget (runs the algorithm with a larger value of $\epsilon$), the data practitioners should gain more accurate information from the data. With DP, $\epsilon$ formally parameterizes the trade-off between accuracy and privacy-loss. For a given algorithm, greater accuracy leads to less privacy guarantee due to information being ``leaked'' from the data. Inversely, the data practitioner could spend a smaller amount of the privacy-loss budget, which will result in less accurate information and greater privacy protection. The goal in devising differentially private algorithms is to optimize this trade-off in order to offer the best accuracy for a given amount of privacy-loss.

%Although DP has existed for over a decade, there is no general consensus on what value of $\epsilon$ should be used for practical implementation. Most data privacy and confidentiality researchers consider the decision a social and policy question that will need stakeholders to shoulder the responsibility of the decision. In addition, the opinions of those involved in creating the DP algorithms (e.g., statisticians or computer scientists) and the participants in the data will have their own opinions on what an appropriate privacy-loss budget is. A full discussion on this topic is beyond the scope of this paper, but we will examine the affects of $\epsilon$ on accuracy within the context of the NIST PSCR Data Challenge.

Many DP algorithms have multiple outputs, such as multiple synthetic data sets or repeated responses from a query system. Each time a statistic or output is released, data information ``leaks'', and therefore needs protecting. DP protects the information by splitting the amount of $\epsilon$ used for each output, and the composition theorems formalize this idea.

\begin{thm}\label{thm:comp} \textbf{Composition Theorems} \citep{mcsherry2009privacy}:
Suppose a mechanism, $\M$, provides $(\epsilon_j$, $\delta_j)$-DP for $j=1,\ldots,k$.
  \begin{itemize}\setlength{\itemindent}{15pt}
  \item[a)] \textbf{Sequential Composition:}\\
    The sequence of $\M_j(X)$ applied on the same $X$ provides $(\sum_j\epsilon_j,\sum_j\delta_j)$-DP.
  \item[b)] \textbf{Parallel Composition:}\\
    Let  $D_j$ be disjoint subsets of the input domain $D$. The sequence of $\M_j(X\cap D_j)$ provides $\max(\epsilon_j), \max(\delta_j)$-DP.
  \end{itemize}
\end{thm}

To put it more simply, suppose there are $k$ many statistical queries on $X$. The composition theorems state that the data practitioner may allocate a portion of the overall desired level of $\epsilon$ to each statistic by sequential composition. A typical appropriation is dividing $\epsilon$ equally by $k$. For example, competitors used  sequential composition in the challenge when making multiple draws of the statistic of interest to generate multiple differentially private synthetic data sets, allocating an equal amount of privacy-budget to each synthetic data set. 

Conversely, parallel composition does not requiring splitting the budget because the noise is applied to disjoint subsets of the input domain. Some competitors used parallel composition, for instance, in a perturbed histogram mechanism, where the bins are disjoint subsets of the data, and noise can be added to each bin independently without needing to split $\epsilon$.

The post-processing theorem is another important theorem, which states that any function applied to a differentially private output is also differentially private.

\begin{thm}\label{thm:post} \textbf{Post-Processing Theorem} \citep{dwork2006calibrating,nissim2007smooth}:

If $\M$ be a mechanism that satisfies $\epsilon$-DP, and $g$ be any function, then $g\left(\M(X)\right)$ also satisfies $\epsilon$-DP.
\end{thm}

Almost all differentially private synthetic data algorithms leverage this theorem, since most focus on perturbing the distribution (either directly or through its parameters) and sampling synthetic data from the noisy distribution. Using the post-processing theorem, any data drawn as a function of the noisy parameters (or a noisy histogram) that were produced by an algorithm satisfying DP will also be DP.
%Additionally, these algorithms tend to incorporate a ``post-processing step'', which becomes vital to improve the accuracy of the final output. 
Other examples of post-processing steps include enforcing structural aspects of the data, such as not releasing negative values for people's ages. Every contestants' algorithm, described in more detail in Section \ref{sec:DPDS}, utilizes some form of post-processing.

\subsection{Differentially Private Mechanisms}

%Algorithms which add sufficient noise to the released data or queries such that they satisfy the differential privacy definitions are commonly referred to as mechanisms. 
In this section, we present some building-block mechanisms that are used in the $\epsilon$-DP and $(\epsilon, \delta)$-DP algorithms developed by the competitors for the NIST PSCR Data Challenge. For a given value of $\epsilon$, an algorithm that satisfies DP or approximate DP will adjust the amount of noise added to the data based on the maximum possible change, given two databases that differ by one row, of the statistic or data that the data practitioner wants released. This value is commonly referred to as the global sensitivity (GS), given in Definition \ref{def:gs}.

\begin{defn}\label{def:gs} \textbf{$l_1$-Global Sensitivity} \citep{dwork2006calibrating}:
For all $X,X'$ such that $d(X,X')=1$, the global sensitivity of a function $u$ is
    \begin{equation}\label{eqn:gs}
        \Delta_1 u= \underset{d(X,X')=1}{\text{sup}} \|u(X)-u(X') \|_1 
    \end{equation}
\end{defn}

We can calculate sensitivity under different norms, for example $\Delta_2 u$ represents the $l_2$ norm global sensitivity, $l_2$-GS, of the function $u$. Another way of thinking about the GS is that it measures how robust the statistical query is to outliers.

The most basic mechanism satisfying $\epsilon$-DP is the Laplace Mechanism, given in Definition \ref{def:lap}, first introduced by \cite{dwork2006calibrating}.

\begin{defn}\textbf{Laplace Mechanism} \citep{dwork2006calibrating}: \label{def:lap}
The Laplace Mechanism satisfies $\epsilon$-DP by adding noise to $u$ that are drawn from a Laplace distribution with the location parameter at 0 and scale parameter of $\Delta_u\epsilon^{-1}$ such that 
    \begin{equation}\label{eqn:lap}
        u^\ast(X)=u(X)+Laplace\left(0,\Delta_1 u \epsilon^{-1}\right)
    \end{equation}
\end{defn}

%Mean zero noise is added to $u(X)$, and the variance of that noise is increased as $\epsilon$ decreases or $\Delta_1 u$ increases. In other words, more noise is added for using a smaller privacy-loss budget or for releasing information that is less robust to outliers. 
Another popular mechanism is the Gaussian Mechanism that satisfies $(\epsilon, \delta)$-DP, given in Definition \ref{def:gauss}, which uses the $l_2$-GS of the statistical query.

\begin{defn}\label{def:gauss} \textbf{Gaussian Mechanism} \citep{dwork2014algorithmic}:
    The Gaussian Mechanism satisfies $(\epsilon,\delta)$-DP by adding Gaussian noise with zero mean and variance, $\sigma^2$, such that
        \begin{equation}\label{eqn:gauss}
            u^\ast(X)=u(X)+N\left(0, \sigma^2 I \right)
        \end{equation}
    where $\sigma=\Delta_2 u \epsilon^{-1} \sqrt{2 \log(1.25/\delta)}$. 
\end{defn}

Both the Laplace and Gaussian Mechanisms are simple and quick to implement, but only apply to numerical values (without additional post-processing, Theorem \ref{thm:post}). A more general $\epsilon$-DP mechanism is the Exponential Mechanism, given in Definition \ref{def:exp}, which allows for the sampling of values from a noisy distribution rather than adding noise directly. Although the Exponential Mechanism can apply to any type of statistic, many theoretical algorithms using the Exponential Mechanism are computationally infeasible for practical applications without limiting the possible outputs for a particular statistic, $\theta$, on $X$. None of the top ranking participants used this mechanism, but other DP synthetic data algorithms such as those proposed by \cite{wasserman2010statistical} and \cite{snoke2018pmse} use the Exponential Mechanism.

\begin{defn}\label{def:exp} \textbf{Exponential Mechanism} \citep{mcsherry2007mechanism}:
    The Exponential mechanism releases values with a probability proportional to
        \begin{equation}\label{eqn:exp}
            \exp \left(\frac{\epsilon u(X, \theta)}{2\Delta_1 u}\right)
        \end{equation}
    and satisfies $\epsilon$-DP, where $u(X,\theta)$ is the score or quality function that determines the values for each possible output, $\theta$, on $X$.
\end{defn}

\section{Differentially Private Data Synthesis Algorithms}\label{sec:DPDS}

In this section, we review the top ranking differentially private data synthesis algorithms from the NIST PSCR Data Challenge. \cite{hay2016principled} and \cite{BowLiu2020} also offer in-depth evaluations and assessments of other differentially private data synthesis methods not covered in this paper, so we direct any interested readers to these papers for more information on algorithms not found here.

This competition, sponsored by the NIST PSCR Division, called for researchers to develop practical and viable differentially private data synthesis methods that were then scored using bespoke metrics. The NIST PSCR challenge consisted of three ``Marathon Matches,'' which spanned from November 2018 to May 2019. Each match provided the contestants with a real-world data set to train and develop their DP methods that were identical in structure and variables to the real-world test data used for final scoring. At the start of each match, organizers of the challenge gave contestants details regarding scoring methods and 30 days to develop and submit their differentially private synthetic data algorithms. The competition required detailed proofs and code for the submissions, and the highest scoring submissions received cash prizes. Over the 30 day period, a panel of subject matter experts reviewed and verified that the submitted methods satisfied DP. If approved, NIST PSCR challenge organizers applied the differentially private synthetic data methods to the test data for final scoring. 

%For the NIST PSCR Data Challenge, competitors were given a public data set to help develop their algorithms before submitting their code for final scoring on the test data. The public and challenge data had identical data structures and variables, so the code could be adequately prepared to run without error. 
Both Matches \#1 and \#2 used the San Francisco Fire Department’s (SFFDs) Call for Service data, with a different year for each match. These data sets contained a total of 32 categorical and continuous variables with roughly 236,000 to 314,000 observations respectively. Some of the variables are \textit{Call Type Group}, \textit{Number of Alarms}, \textit{City}, \textit{Zip Code of Incident}, \textit{Neighborhood}, \textit{Emergency Call Received Date and Time}, \textit{Emergency Call Response Date and Time}, \textit{Supervisor District}, and \textit{Station Area}. For Match \#3, challenge participants trained their methods on the Colorado Public Use Microdata Sample (PUMS) data, and their methods were evaluated on the Arizona and Vermont PUMS data for final scoring. All three PUMS data sets had 98 categorical and continuous variables with the number of observations ranging from about 210,000 to 662,000. \textit{Gender}, \textit{Race}, \textit{Age}, \textit{City}, \textit{City Population}, \textit{School}, \textit{Veteran Status}, \textit{Rent}, and \textit{Income Wage} were a few of the 98 variables. We discuss how the NIST PSCR Differential Privacy Synthetic Data Challenge executed their scoring in Section \ref{sec:DP-Metric}.

We categorize the differentially private data synthesis methods from the challenge into the same two categories used in \cite{BowLiu2020}, non-parametric and parametric approaches. We define non-parametric approaches as differentially private data synthesis methods that generate data from an empirical distribution, and we define parametric approaches as algorithms that generate the synthetic data from a parameterized distribution or generative model.

\subsection{Non-Parametric Data Synthesis}\label{subsec:DP-Hist}

Most non-parametric differentially private synthetic data techniques sanitize the cell counts or proportions from a cross-tabulation of the data. 
%(e.g., the full cross-tabulation of all variables) For categorical data, the noisy tables can be released. 
The non-parametric approaches will sample data from an empirical distribution using the discretized bins to provide a synthetic microdata file or when the original data has continuous variables. The bounds for the discretization of continuous variables must be selected in a differentially private manner or by leveraging public information to satisfy DP. 
%which is often a tricky part of these methods. 
The majority of the teams who developed non-parametric data synthesis methods focused on reducing the number of cells to sanitize. They accomplished this by clustering variables (i.e., creating multiple disjoint cross-tabulations on subsets of variables), maintaining only highly correlated marginals, or using the privacy budget asymmetrically across cells.

%consistent opinion regarding easy of use/understanding, computational burden, and former acceptance in the literature
\subsubsection{Team DPSyn}\label{subsub:dpsyn}
Team DPSyn consistently performed well throughout the entire NIST PSCR Data Challenge, placing second in all three matches. The team's mechanism, \textit{DPSyn}, works by clustering similar variables (based on attributes, the specific utility objective, etc.) and perturbing the cell counts of the joint histograms for each cluster. Team DPSyn used the training data set in order to determine clusters of variables. Because this was done using public data rather than the sensitive values, they did not need to spend any additional privacy budget. Overall, the approach lessens the noise necessary, because it reduces the total number of cells, but at the price of sacrificing the correlations between variables in different clusters. After clustering, \textit{DPSyn} constructs the 1-, 2-, and 3-way marginals for all variables in each cluster, and sanitizes the counts via the Gaussian Mechanism. For post-processing, \textit{DPSyn} constrains the noisy marginals using techniques from \cite{qardaji2014priview} to be consistent with one another. These techniques check for mutual consistency among the totals of the multi-way marginals (altering the counts to be consistent) and reduce the noisy counts to zero when they are below a threshold. Finally, \textit{DPSyn} generates the synthetic data by sampling from the noisy marginals of the joined clusters.

The algorithm is straightforward, and a data practitioner could implement \textit{DPSyn} fairly easily given the simplicity and because \cite{li2019} provided the source code (\verb;Python;) and full documentation on GitHub. The main difficulty would be selecting the variable groups for the pre-processing step, which could be daunting for an inexperienced data practitioner, someone without familiarity of the data set, or someone who does not have access to public data. This method is fairly novel, only being published recently, so it has yet to gain wide acceptance in the field. That being said, more researchers and data practitioners will likely implement \textit{DPSyn} in the near future due to its simplicity and good performance.

\subsubsection{Team Gardn999}\label{subsub:gardner}

\begin{comment}
\begin{algorithm}[!htb]
\caption{DPFieldGroups}\label{alg:gardner}
    \textbf{Input:} target data set: $X$ with dimensions ${n\times q}$, public data set: $Y$ with dimensions ${m\times q}$, privacy-loss budget: $\epsilon$
    \begin{algorithmic}[1]
        \State Identify and group highly correlated variables in $X$ based on the publicly available data set, $Y$. Let $k\leq q$ represent the total number of groups identified.
        \State Create histograms of the grouped variables such that $B_j$ (for $j=1,...,k$) is the total number of all possible variable (field) combinations.
        \State Sanitize the counts, $\theta_{j,l}$ (for $l=1,...,B_j$), in each histogram bin such that $\theta^*_{j,l}=\theta_{j,l}+e_{j,l}$, where $e_{j,l}\sim Lap(k/\epsilon)$ and $\theta^*_{j,l}$ is rounded to the nearest non-negative integer.
        \State Set $\theta^*_{j,l}=0$ for all $\theta^*_{j,l}\leq k/\epsilon\cdot \log_{10}(B_j)$.
        \For{$i=1$ to $n$} 
            \State Select $\theta^*_{j,l}$ with a probability proportional to $\theta^*_{j,l}/\sum^{B_j}_{l=1} \theta^*_{j,l}$ for all $j=1,...,k$.
            \State Generate a single observation of the synthetic data, $\{\x_i\}=X^*$, from the variable attribute values that correspond to the selected $\theta^*_{j,l}$.
        \EndFor
    \end{algorithmic}
    \textbf{Output:} synthetic data set: $X^*$
\end{algorithm}
\end{comment}

Team Gardn999 developed the simplest mechanism, \textit{DPFieldGroups}, out of the NIST PSCR Data Challenge entrants while still performing well. They placed fifth and fourth in Matches \#2 and \#3, respectively, and did not participate in Match \#1. \textit{DPFieldGroups} sanitizes the original data cell counts via the Laplace Mechanism. \textit{DPFieldGroups} first clusters the cells by identifying the highly correlated variables from the public training data set. The method then conducts post-processing by reducing noisy counts to zero if they fell below a threshold calculated from $\epsilon$ and the $\log_{10}$ number of bins in the particular marginal histogram. \textit{DPFieldGroups} generates the synthetic data by randomly sampling the sanitized observations from each of the marginal histograms with a weighted probability proportional to the noisy counts. 

Similarly to \textit{DPSyn}, the pre-processing step for \textit{DPFieldGroups} relies on the data practitioner to cluster highly correlated variables based on public data. Once the variables are grouped, the data practitioner can execute the \verb;Java; code hosted on GitHub \citep{gardn999}. The post-processing step is less involved than \textit{DPSyn}, only adjusting some counts down to avoid a large number of non-zero bins. The strength of this approach lies in its simplicity. On the other hand, Team Gardn999 has not published \textit{DPFieldGroups} as a novel method, and it relies only on relatively simple DP steps. However, this method forms a good case study for the performance of a simple application of a differentially private algorithm.

\subsubsection{Team pfr}\label{subsub:pfr}

Team pfr placed first in Matches \#1 and \#2, but did not compete in Match \#3. We believe their lack of participation might be due to their initially designing their algorithm based on how Match \#1 scored the similarity of the original and synthetic data sets, and they did not want to recalibrate their algorithm for the new data and scoring metric in Match \#3. They targeted maximizing accuracy on the 3-way marginal counts, and the variables that involve the Emergency Call Data and Time information. Before sanitizing the 3-way marginals, their pre-processing step depends on the data practitioner establishing a list of:
\begin{enumerate}[topsep=0pt,itemsep=-1ex,partopsep=1ex,parsep=1ex]
    \item variables that could be computed deterministically from other variables and therefore did not need to be encoded, e.g., \textit{City} was computed deterministically from \textit{Neighborhood}.
    \item variables that are correlated or variables that are subset of others. e.g., \textit{Supervisor District} is correlated with \textit{Station Area}.
    \item data set size thresholds for certain queries, such that queries over the threshold are discarded and the corresponding output is replaced by a uniform distribution.
\end{enumerate}
Similarly to \textit{DPSyn}, Team pfr used the training data set as their public data to determine the histogram queries and the deterministic relationships. The algorithm roughly clusters the variables into three disjoint groups (Spatial, Temporal, and Call-Information groups). The \textit{pfr} method then identifies within each group which variables are computed deterministically from other variables and which variables are highly correlated. Clustering the variables in this manner reduces the total possible combinations of cells that need sanitizing. For the sanitizing step, the \textit{pfr} mechanism sanitizes the cell counts in each group of variables separately via the Laplace mechanism. Team pfr allocates the privacy budget proportionally to the number of variables in each group. e.g., \textit{Call-Information} had 10 variables out of 32 possible, so a total of $\approx 0.31\epsilon$ privacy budget. After sanitizing the counts within each group, Team pfr's approach ``denoises'' or lowers the total amount of noise added to the counts by modeling all the non-negative counts from a mixture probability distribution. This distribution samples either a zero or values from a uniform distribution to reduce the excessive non-zero counts created in the sanitization step. Team pfr uses the public training data set to estimate the probability of sampling a zero count based on the proportion of empty cells within each grouping.
%Finally, the mechanism normalizes to the desired synthetic data set size.

A data practitioner would have to hand-code the \textit{pfr} algorithm given the lack of open source code (the team did not share their code on GitHub) or existing publication of the approach. Without public code or an algorithm, their method is not reproducible. Also, the \textit{pfr} method depends heavily on the publicly available information for query selection to improve accuracy, so this method would likely perform poorly on data sets with little to no associated public knowledge. Team pfr may not publish or receive credit in the literature for their ideas, but they demonstrated how simpler DP methods that intelligently leverage public or domain knowledge can perform well in practice.

\subsection{Parametric Data Synthesis}\label{subsec:DP-Mod}

Parametric differentially private synthetic data methods rely on estimating or learning an appropriate parameterized distribution based on the original data and sampling values from that distribution with noisy parameters. Parametric methods are generally much more computationally demanding than the non-parametric methods. One of the concerns when applying a parametric approach is the distribution or model selection itself might violate privacy. Either the data practitioner must use a separate public data set to test what model is appropriate or leverage public knowledge on what model should be used to avoid a privacy violation. If this is not possible, the data practitioner may apply a differentially private model selection method \citep{lei2018differentially}.

\subsubsection{Team PrivBayes}\label{subsub:privbayes}
Team PrivBayes used a well developed differentially private approach, \textit{PrivBayes}, from their well-cited paper \citep{zhang2017privbayes}. They placed fifth in Match \#1 and third in Matches \#2 and \#3. \textit{PrivBayes} uses a Bayesian network with binary nodes and low-order interactions among the nodes to release high-dimensional data that satisfies DP. \textit{PrivBayes} first scores each pair of possible attributes that indicates the level of correlation between attributes. The method sanitizes these scores via the Gaussian Mechanism, and then uses them to create the Bayesian network. When the attributes contain continuous values, \textit{PrivBayes} must discretize the values to create the Bayesian network. Using this differentially private Bayesian network, the algorithm approximates the distribution of the original data with a set of $P$ many low-dimensional marginals. Next, \textit{PrivBayes} sanitizes the $P$ marginals via the Gaussian Mechanism and uses the noisy marginals with the Bayesian network to reconstruct an approximate distribution of the original data set. \textit{PrivBayes} then generates the synthetic data by sampling tuples from this approximated distribution, and post-processes to enforce consistency on the noisy marginals in three parts: marginal set of attributes, attribute hierarchy, and overall consistency. In other words, this method first checks that the marginal counts are consistent for a chosen set of attributes, then enforces that the marginals remain consistent as the attribute coarsens. For example, the set of attributes could be ethnicity within census tracts for a state, where the noisy marginal counts for each ethnic group should sum to the noisy total count of individuals for the particular census tract. For attribute hierarchy consistency, the sum of certain census tracts should equal the total counts for county and the state. For overall consistency, all marginals for all subsets of attributes should be consistent.

\textit{PrivBayes} performs fairly well and does not require any public data for a pre-processing step such as the non-parametric approaches described in Section \ref{subsec:DP-Hist}. Additionally, there exists \textit{PrivBayes} \verb;Python; code on GitHub \citep{DataSynthesizercodes,DataSynthesizer}, allowing data practitioners to easily apply \textit{PrivBayes} to their data. However, the complexity of \textit{PrivBayes} due to constructing the differentially private Bayesian network and enforcing consistency among the noisy marginals increases the computational burden compared to the other methods. A data practitioner might be limited in implementing \textit{PrivBayes} depending on computational resources and the size of the target data set. The complexity of the approach, notably the unsupervised identification of the network, also means that data practitioners will have more difficulties diagnosing potential issues in the event of inaccurate syntheses. While the non-parametric methods are easy to understand and tune, \textit{PrivBayes} essentially represents a black-box method. That being said, researchers and practitioners may find the well founded theory and acceptance of \textit{PrivBayes} in the literature a boon to its potential use.

\subsubsection{Team RMcKenna}\label{subsub:rmckenna}
Team RMcKenna performed third in Match \#1, fourth in Match \#2, and first in Match \#3. Their approach can be described as a blend of a parametric and a non-parametric approach, since the algorithm focuses on determining a subset of histogram cells to perturb and then sampling data from these noisy marginals using a graphical model. As a first step, Team RMcKenna's mechanism uses a similar pre-processing step to the non-parametric methods by first identifying the highly correlated variables on a public data set. The algorithm then sanitizes  the 1-, 2-, and 3-way marginals via the Gaussian Mechanism. In this step, the team also utilizes the Moments Accountant, a privacy-loss tracking technique that tightens the privacy bound for the Gaussian Mechanism better than Theorem \ref{thm:comp}, resulting in less noise on the marginals \citep{abadi2016deep}. Based on the sanitized marginals, Team RMcKenna uses graphical models to determine a model for the data distribution, capturing the relationships among the variables and enabling synthetic data generation \citep{mckenna2019graphical}.

Team RMcKenna's method is fairly easy to understand, resembling the implementation steps for \textit{DPSyn} and \textit{DPFieldGroups}, while utilizing some more advance techniques for splitting the privacy budget across cells and sampling from the noisy marginals. The combination of the parametric and non-parametric ideas offers a unique approach among the competitors. The algorithm is also straightforward to implement, requiring only some pre-processing work. The data practitioner must first select the highly correlated variables for the low dimensional marginals before executing the \verb;Python; code from \cite{rmckenna} on GitHub. This method is fairly novel and, given its performance and the fact that it builds on previous work, it will likely gain acceptance in the literature.

\subsubsection{Team UCLANESL}\label{subsub:ucla}

Team UCLANESL placed fourth in Match \#1, fifth in Match \#3, and they did not compete in Match \#2. They based their mechanism on the Wasserstein generative adversarial network (WGAN) training algorithm along with the Gaussian Mechanism and the Moment Accountant technique to ensure DP \citep{arjovsky2017wasserstein}. First, WGAN trains two competing models: the \textit{generator}, a model that learns to generate synthetic data from the target data, and the \textit{discriminator}, a model that attempts to differentiate between observations from the training data and the \textit{generator} created synthetic data. The \textit{generator} creates fake observations that mimic ones from the target data by taking in a noisy vector sampled from a prior distribution such as a normal or uniform distribution. These fake observations attempt to confuse the \textit{discriminator}, reducing the model's ability to distinguish the target and synthetic data sets. For the models to be differentially private, Team UCLANESL's method sanitizes the \textit{discriminator} gradient updates using the Gaussian mechanism. Essentially, the method first  ``clips'' the \textit{discriminator} gradient updates to ensure a bounded $l_2$ sensitivity before adding noise from the Gaussian Mechanism. Team UCLANESL's approach then uses these sanitized gradient updates on the \textit{discriminator} model weights, which means the \textit{generator} model also satisfies DP since it relies on the feedback from the \textit{discriminator}. The Moment Account technique comes in to track the privacy-loss budget and will abort the WGAN training if the privacy budget has been reached.

For further details, \cite{uclanesl_dp_wgan} provides a full technical report with proofs in addition to their \verb;Python; code. As a published paper with publicly available code, data practitioners could easily implement this method. However, Team UCLANESL's method is the most computationally intense out of all the competitors. In particular, their method consumes a lot of memory. This in large part due to the computational nature of GANs. A second team also submitted a GAN algorithm for the challenge, but the NIST PSCR competition staff could not even get the code to run. Team UCLANESL's \verb;Python; code includes the TensorFlow library, a GPU-accelerated deep learning framework, that they report significantly reduces the computational time when the code runs on a GPU-powered machine. For this reason, we suspect the average data provider, who would likely not have access to a GPU, will have extreme difficulties implementing the DP WGAN method given the computational resources required.

\subsection{Summary of the NIST PSCR Challenge Synthesis Algorithms}

In this section, we offer our high-level evaluation of the contestants' algorithms based on their theoretical strengths and weaknesses along with their commonalities and dissimilarities. We also consider the relative applicability for a practitioner wishing to release data based on the required pre-processing and computational demands of each algorithm. Table \ref{tab:DIPS} provides summaries for each algorithm.

\begin{table}[htbp]
    \def\arraystretch{1.1}
    \caption{Summary of the non-parametric and parametric differentially private synthetic data approaches discussed in Sections \ref{subsec:DP-Hist} and \ref{subsec:DP-Mod}.}
    \label{tab:DIPS}
    \centering\small
    \begin{tabular}{ C{3.0cm} | L{3cm} | L{3.5cm} | L{5cm}}
    \hline
        \multicolumn{4}{c }{\textit{Non-Parametric Synthesis Approaches}} \\
    \hline
        \textbf{Team} & \textbf{Computation}  & \textbf{Off-the-Shelf vs. Hand-Coding} & \textbf{Pre- and Post-Processing} \\
    \hline
          \makecell{Team DPSyn \\ (Sec. \ref{subsub:dpsyn})} & light to moderate computational complexity & some hand-coding due to identifying marginals for pre-processing, \verb;Python; code available on GitHub & \textbf{pre-processing:} identify marginals from public data; \textbf{post-processing:}  adjust noisy marginals to be consistent and change counts to zero below a threshold\\
    \hline
          \makecell{Team Gardn999 \\ (Sec. \ref{subsub:gardner})} & simplest and fastest method & some hand-coding due to identifying marginals for pre-processing, \verb;Java; code available on GitHub & \textbf{pre-processing:} identify marginals from public data; \textbf{post-processing:} adjust the overall counts based on a threshold to avoid a large number of non-zero bins\\
    \hline
         \makecell{Team pfr \\ (Sec. \ref{subsub:pfr})} & simple and quick after pre-processing & hand-coding required, no public code available & \textbf{pre-processing:} identify marginals from public data; \textbf{post-processing:} reduce the number of non-empty cells from sanitization by modeling the noisy cell counts\\
    \hline
        \multicolumn{4}{c }{\textit{Parametric Synthesis Approaches}} \\
    \hline
        \textbf{Team} & \textbf{Computation}  & \textbf{Off-the-Shelf vs. Hand-Coding} & \textbf{Pre- and Post-Processing} \\
    \hline
        \makecell{Team PrivBayes \\ (Sec. \ref{subsub:privbayes})} & more computationally complex compared to the other methods & off-the-shelf via \verb;Python; code on GitHub & \textbf{pre-processing:} automated Bayesian network to determine which variables are highly correlated or not; \textbf{post-processing:} enforcing consistency among the marginals\\
    \hline
         \makecell{Team RMcKenna \\ (Sec. \ref{subsub:rmckenna})} & light to moderate computational complexity & some hand-coding due to identifying marginals for pre-processing, \verb;Python; code on GitHub & \textbf{pre-processing:} identify marginals from public data\\
    \hline
        \makecell{Team UCLANESL \\ (Sec. \ref{subsub:ucla})} & the most computationally complex method; requires more RAM memory & off-the-shelf via \verb;Python; code on GitHub & none \\
    \hline
    \end{tabular}
\end{table}

The three non-parametric algorithms function similarly, relying on estimating histograms on a reduced numbers of cells and perturbing the counts. These methods then draw the synthetic values from these noisy marginals. The differences come from how they construct the histograms, how they allocate the privacy budget, and what post-processing they use. In fact, teams DPSyn and Gardn999 have almost the exact same core approach, except with \textit{DPSyn} offering additional pre- and post-processing techniques. In contrast, the parametric approaches vary significantly from one another. Team PrivBayes relies on Bayesian networks, Team UCLANESL uses a GAN technique, and Team RMcKenna layers a graphical model on top of a perturbed histogram. These methods highlight the fact that non-parametric algorithms require significant hands-on work apart from the actual privacy mechanism, while parametric algorithms focus on optimizing the privacy mechanism itself.

The non-parametric methods are much less computationally demanding, but they require more pre-processing work such as analyzing public data to identify correlations and important marginals. For a data provider wishing to implement one of these methods, they will need to spend time working with public data and likely perform some hand-coding. Only one method, \textit{PrivBayes}, truly qualifies as an ``off-the-shelf" method, where it requires no prep work or additional coding to run it. Although Team UCLANESL has open-source code and does not require additional coding, the approach demands such significant computational resources, namely a GPU, that we expect few practitioners could run it in practice without changes to their computational environment. The other four teams' algorithms (DPSyn, Gardn999, pfr, and RMcKenna) need detailed pre-processing before running the code. These four methods assume access to accurate public data, so practitioners without such available information would not benefit from implementing one of these algorithms.

Based on the descriptions of the methods, we expect data practitioners would most easily adopt \textit{DPSyn} (or to a lesser extent Team Gardn999's \textit{DPFieldGroups}) and \textit{PrivBayes}. The former requires little computational or technical understanding, but involves some effort with analyzing the public information beforehand. \textit{PrivBayes} on the other hand is truly off-the-shelf and could be applied without pre-processing based on public data, assuming the practitioner has the required computational abilities. In contrast, teams pfr, RMcKenna, and UCLANESL offer more complex approaches that may provide good results for more expert users.

\section{Metrics to Evaluate the Synthetic Data Quality}\label{sec:DP-Metric}
In this section, we describe the scoring methods used for the NIST PSCR Differential Privacy Synthetic Data Challenge. We also detail the quality metrics we used for further evaluation of the DP synthetic data sets, which includes general joint distributional level measures, marginal distributional differences, and differences in specific fitted regression models.

\subsection{NIST PSCR Differential Privacy Synthetic Data Challenge Scoring}\label{subsec:nistmetric}

\begin{table}[!htb]
    \def\arraystretch{1.2}
    \caption{NIST PSCR Differential Privacy Synthetic Data Challenge Marathon Match Information.}
    \label{tab:NIST-match}
    \centering\small
    \begin{tabular}{ C{1.2cm} | C{4.20cm} | C{4.20cm} | C{4.20cm}}
    \hline
        \textbf{Match} & \textbf{Training Data}  & \textbf{Scoring Data} & \textbf{Analyses} \\
    \hline
          1 & 2017 SFFD's Call for Service Data & 2016 SFFD's Call for Service Data & ``Clustering" \\
    \hline
          2 & 2016 SFFD's Call for Service Data & 2006, 2017 SFFD’s Call for Service Data & ``Clustering" and ``Classification" \\
    \hline
          3 & Colorado PUMS & Arizona and Vermont PUMS & ``Clustering", ``Classification", and ``Regression" \\
    \hline
    \end{tabular}
\end{table}

We summarize the scoring analyses used for the three ``Marathon Matches'' from the NIST PSCR Differential Privacy Synthetic Data Challenge in Table \ref{tab:NIST-match}. For each match, the final scores were progressively evaluated based on bespoke metrics termed ``clustering", ``classification", and ``regression". This means Match \#1 had only the clustering analysis, Match \#2 had the clustering and classification analyses, and Match \# 3 used all three analyses. NIST PSCR announced the metric criteria at the start of each match, so the competitors could modify their approach based on the scoring metrics. The ``clustering" analysis compared the 3-way marginal density distributions between the original and synthetic data sets, where the utility score was the absolute difference in the density distributions. NIST PSCR repeated this calculation 100 times on randomly selected variables, and then averaged for the final clustering score. The ``classification" analysis first randomly picked 33\% of the variables. If a particular variable was categorical, the method randomly picked a subset of the possible variable values, whereas, if the variable was continuous, it randomly picked a range of values. The analysis then used these selected values to calculate how many of the observations in the synthetic and original data matched the specific variable subset. Finally, the synthetic data counts were subtracted from the original data matched counts before taking the natural log. NIST PSCR computed the natural log difference over 300 repeats, and the final classification score was the root mean-squared on the repetitions divided by $ln(10^{-3})$. The term ``classification" is slightly misleading, given that this was essentially testing similarity between the original and synthetic data in the randomly selected subsets of the joint distributions. Lastly, the ``regression" analysis used a two-part score system. The first score calculated the mean-square deviation of the Gini indices in the original and synthetic data sets for every city on the gender wage gap, and then averaged those values over the total number of cities in the original data. The second score compared how the cities in the original and the synthetic data sets were ranked on gender pay gap, calculating the rank differences by the mean-square deviation. NIST PSCR averaged these two scores for the overall regression analysis score. Again, the term regression is slightly misleading given that this was not a comparison of regression coefficients, as is commonly seen in literature. For all three NIST PSCR scoring metrics, a larger value in the challenge indicated that the synthetic data preserved the original data well. Note that in section \ref{sec:utilty_eval} when we present our full utility evaluation, we rescale the NIST scores, such that a smaller value indicates high utility.

\subsection{General Discriminant-based Quality Metric Algorithms}\label{subsec:metric}

We now describe the general utility approaches to measure overall distribution similarity between synthetic and the original data that we employed in our extended evaluation. These metrics should give a broad sense of how ``close'' the synthetic data are to the original data. The approaches presented here utilize the concept of propensity scores (predicted probabilities of group membership) to discriminate between the original and synthetic data, and the corresponding utility metrics are calculated in different ways using the estimated propensity scores. Researchers first developed these methods on traditional synthetic data, but they apply to differentially private synthetic data as well. At a high level, these utility measures train a classifier to discriminate between two data sets, and the more poorly a classifier performs, the more similar the data sets are assumed to be on a distributional level.

\cite{woo2009global} first proposed using propensity scores and summarized them into a utility metric by calculating the mean-square difference between the propensity score estimates and the true proportion the synthetic data within the total combined data set. \cite{snoke2018general} later coined the value as the propensity score mean-squared error (\textit{pMSE}) and improved the \textit{pMSE} by deriving its theoretical expected value and standard deviation under certain null conditions. The authors used these values to create standardized versions of the statistic called the \textit{pMSE}-ratio and standardized \textit{pMSE}. \cite{sakshaug2010synthetic} applied a Chi-squared test on the discretized estimated propensity scores. \cite{bowen2018statistical} developed SPECKS (\textbf{S}ynthetic data generation; \textbf{P}ropensity score matching; \textbf{E}mpirical \textbf{C}omparison via the \textbf{K}olmogorov-\textbf{S}mirnov distance), which applies the Kolmogorov-Smirnov (KS) distance to the predicted probabilities as the utility metric. A small KS distance indicates that the original and synthetic empirical CDFs are indistinguishable.

To produce the results in Section \ref{sec:study}, we calculate the \textit{pMSE}-ratio and SPECKS. Both approaches require training and fitting classifiers to the combined original and synthetic data with a binary indicator labeling the data set in each row. We obtain the predicted probabilities of this binary label, and we compute the \textit{pMSE} using
\begin{equation}
    pMSE = \frac{1}{N}\sum^N_{i=1} (\hat{p}_i-c)^2
\end{equation}
where $N=n+n^*$ is the total number of observations from both the original and synthetic data and $c=n^*/N$ is the proportion of observations from the synthetic data out of the total. The value of $c$ is often 0.5 because synthetic data is typically generated with the same number of rows as the original data, but this constraint was not made for the NIST PSCR challenge. When we use a parametric model for the classifier, \cite{snoke2018general} derived theorems for the expected value under the null hypothesis that the original and synthetic data were sampled from the same generative distribution. We calculate the expected null mean using
\begin{equation}\label{eqn:pmse-exp}
    \mathbb{E}(pMSE)=(k-1)\frac{(1-c)^2c}{N}
\end{equation}
such that $k$ is the number of parameters from the classifier. If we use a non-parametric classifier, we can approximate the null expected value using resampling techniques such as permutating the rows and reestimating the \textit{pMSE}. After calculating the null mean \textit{pMSE}, we obtain the \textit{pMSE}-ratio by dividing the observed mean \textit{pMSE} calculated on the original and synthetic data by the null mean. A \textit{pMSE}-ratio value close to 1 indicates that the synthetic data is drawn from a distribution that approximates the generative distribution of the original data. For a more in-depth discussion of this method such as its strengths and weaknesses, please see \cite{snoke2018general}.

We make two changes to the original \textit{pMSE}-ratio as proposed by \cite{snoke2018general}. First, we choose the best classification and regression trees (CART) models for estimating the propensity scores using cross-validation. Because the \textit{pMSE} is sensitive to the classification model, using different levels of complexity in the CART models result in different utility values. To aid our model choice, we set the CART complexity parameter (CP), which controls how large the trees grow, by performing cross-validation and choosing the CP that minimized the error. This ensures our utility does not ``overfit" to the data, i.e., we use a good distributional discriminator between the original \textit{distribution} and the synthetic data. Running cross-validation for each combination of the original data and each of the competitors' data, we found roughly the same best CP value, so we constrained the CP to be the same for all competitors without any additional adjustments. Following the prior work, we must use the same CP for each synthetic data set, because the \textit{pMSE} values are not comparable when using different classifiers. For example, a \textit{pMSE} computed from a logit model with only first-order terms measures a different type of distributional similarity than a \textit{pMSE} computed using first-order and interaction terms. In the same way, \textit{pMSE} calculated from CART models with different CPs measure relationships in the data on different levels of complexity.

Second, and more importantly, we changed how we estimate the null \textit{pMSE} in the CART models. An issue with the pairwise or permutation approach originally proposed is that while these approaches measure the null for two data sets that came from approximately the same distribution, they did not necessarily (and very likely did not) come from the generative distribution of the original data. In theory, this should not matter because the expected value of the \textit{pMSE} under the null depends on the classifier model not the data, but CART models' complexity changes based on the data and the expected value under the null will depend on the complexity of the CART model. This means in practice when we calculate the null value for CART models, the value will be larger if we use synthetic data that comes from a more varied generative distribution. In other words, if the synthetic data sets are further apart, then the null value is larger than if the synthetic data sets are closer together. The permutation or pairwise process produces different estimated nulls for different synthetic data models, even with the same CART complexity parameter, which contradicts the theory that the null stays fixed regardless of the synthetic data. 

The differences in estimated null values may be minimal with non-differentially private synthetic data, but we find they are much more varied for differentially private synthetic data. As $\epsilon$ decreases, the noise in the synthetic data models increases significantly such that the estimated null becomes quite large (because each synthetic data set is very far from each other). This increase cannot be matched by an increase in the observed \textit{pMSE}, because that value is bounded above. In these situations, when using the \textit{pMSE}-ratio, we observe that the algorithm suggests ``better" utility using lower $\epsilon$ than using higher $\epsilon$, because the null increases with lower $\epsilon$ and thus lowering the ratio.
 
In this paper, we solve this problem by instead estimating the null by only using the original data. Instead of calculating the pairwise comparisons or permuting the original data with the synthetic, we bootstrap two times the number of rows from the original data, and we assign 0 labels to half and 1 labels to the other half. We then calculate the \textit{pMSE}, repeat this process 100 times, and take the average value as our null \textit{pMSE}. This approach ensures the null value arises from the original data, which we know came from the observed generative distribution, and the estimated null does not differ for different synthetic sets, which it should not.

For SPECKS, we determine the empirical CDFs of the propensity scores for the original and synthetic data sets separately, and then apply the Kolmogorov-Smirnov (KS) distance on the two empirical CDFs. The KS distance is the maximum distance of two empirical CDFs, where the synthetic and the original data have the largest separation. A smaller KS distance (close to 0) indicates that the synthetic data preserved the original data well whereas a larger KS distance (close to 1) means the synthetic data differs a lot from the original data. For more on this method, please see \cite{bowen2018statistical}.

We can use both of these methods with either simple parametric models, such as logistic regression, or more complex non-parametric models, such as CART. For any given model, we can compare different synthetic data sets using these metrics. One issue comes from the fact that we may obtain different rankings if we use different classifiers. This is because models with varying complexity are actually measuring different types of distributional similarity. A logistic regression with only main effects, for example, only measures the accuracy of the first-order marginal distributions (simultaneously). A more complex CART model on the other hand is measuring high-order interactions in the data. As was recommended in the previous work, we use different classifiers and compare relative rankings from each set of models. This approach gives multiple views of the utility of the data.

\subsection{Marginal Distributional Metrics and Regression Analyses}

Lastly, we describe quality metrics which measure accuracy on more specific elements of the data, such as the univariate differences between the synthetic and the original data or differences in the estimates from regression analyses. DP work frequently uses univariate distance measures to assess the amount of noise added due to a privacy mechanism. Typically in the literature, the measures include the average $l_1$ distance, the root-mean-squared error, or various other ways to measure distance. We choose to use a different approach for a few reasons. First, rather than using a direct distance measure, we utilize a distributional distance metric. These measures make sense for synthetic data, since they measure distributional distance and do not require vectors of the same size. This enables us to compare the original data with various synthetic data sets that have different numbers of observations, which were not constrained for the challenge. The distributional metrics we use are the Chi-square test for categorical variables and the KS test for continuous variables. Lastly, because each variable in the data has a different scale, we average the univariate distances across the whole data set by first converting to p-values. We are not using p-values in the traditional null hypothesis significance testing (NHST) framework, but rather we are just using them as a scale-free distance measure. We could achieve something similar by rescaling all the variables in our data and using the Chi-square and KS test statistics. Our univariate utility metric is then the average of the p-values for each variable. Categorical and continuous variables are treated separately, since one might argue that the processes of synthesizing a continuous or categorical distribution are different.

Previous work in the synthetic data literature also often compares the results from regression models fit on both the original and synthetic data, measuring how much the analyses are effected by the privacy mechanism. We perform two regression models based on \cite{simon2009higher}, who used the 1940 Census data, and we compute two specific utility metrics for each of the coefficients in the models. The first is the confidence interval overlap measure, proposed by \cite{karr2006framework}, among others, which is defined as

\begin{equation}\label{eqn:io}
IO = 0.5 \bigg( \frac{min(u_o, u_s) - max(l_o, l_s)}{u_o - l_o} + \frac{min(u_o, u_s) - max(l_o, l_s)}{u_s - l_s} \bigg)
\end{equation}

\noindent where $u_o$, $l_o$ and $u_s$, $l_s$ are the upper and lower bounds for the original and synthetic confidence intervals respectively. This measures how much the confidence intervals estimated the original and synthetic data overlap for a single estimate on average, where the maximum value is 1. Along with this, we calculate the standardized difference in coefficient values, i.e., $|\hat \beta_o - \hat \beta_s|/se(\hat \beta_o)$, used by \cite{woo2015generalised} and \cite{snoke2018general}. This measures how far the synthetic data coefficients lie from the original quantities instead of considering the width of the confidence interval. A drawback to the IO measure is the inability to distinguish whether the it is the synthetic data or the original data that has a wider confidence interval that encompasses the other interval. The standardized difference, on the other hand, depends only on the point estimate and original coefficient variability. Together, the two metrics allow us to more accurately assess the inferential differences between the original and synthetic data regression models.

\subsection{Utility Metric Categories}

Table \ref{tab:util-categories} defines the three categories of the utility metrics we use to evaluate the synthetic data. The metrics include both the original NIST PSCR metrics and additional metrics. We adjusted the metrics as much as possible to be on the same scale, i.e., $[0, 1]$, though some are unbounded and thus could not be rescaled.

\begin{table}[!htb]
    \def\arraystretch{1.2}
    \caption{Utility Metric Categories.}
    \label{tab:util-categories}
    \centering\small
    \begin{tabular}{ C{4.75cm} | C{4.75cm} | C{4.75cm} }
    \hline
        \textbf{Marginal Distribution Metrics} & \textbf{Joint Distribution Metrics}  & \textbf{Correlation Metrics} \\
    \hline
          Chi-Sq Distance (Categorical Variables) & \textit{pMSE}-ratio & Regression Coefficient Confidence Interval Overlap \\
    %\hline
          KS Distance (Continuous Variables) & SPECKS & Regression Coefficient Standardized Difference \\
    %\hline
          NIST PSCR ``Classification'' Task  & NIST PSCR ``Clustering'' Task & NIST PSCR ``Regression'' Task \\
    \hline
    \end{tabular}
\end{table}

In many ways, the general and specific measures discussed in the previous sections formalize concepts underlying the scoring metrics devised for the NIST PSCR Data Challenge, which also sought to assess distributional similarity or specific analytical similarity. Correspondingly, each of our categories has one of the NIST PSCR measures, even though they are not perfect mappings. For example, the NIST PSCR ``classification" task measures 3-way marginals rather than 1-way marginals. 
%Our goal is to show both how the metrics within these groupings are similar and not exactly the same. 
For data practitioners who wish to evaluate the utility of the DP synthetic data algorithms, the formally developed metrics we use offer two primary advantages over the bespoke metrics used in the challenge. First, they will be easier to implement, since for many of the metrics software already exists to compute them. Second, they have statistical interpretations that are clear and easy to understand.

\section{Experimental Results}\label{sec:study}

We summarize the results from all of the utility evaluations in this section, including the NIST PSCR Differential Privacy Synthetic Data Challenge, the \textit{pMSE}-ratio, the SPECKS metrics, the univariate distribution comparisons, and the regression models. NIST PSCR gave us access to the original synthetic data generated by the competitors listed in Table \ref{tab:NIST-results}. The metrics we present provide a broader picture than those used in the challenge alone, and, from these results, we make further assessments of the different DP synthetic data algorithms.

\subsection{NIST PSCR Differential Privacy Synthetic Data Challenge}\label{subsec:match}

For Matches \#1 and \#2, NIST PSCR set the privacy-loss budget at $\epsilon=\{0.01, 0.1, 1\}$ and $\delta=0.001$ (some contestants chose not to use $\delta$), whereas for Match \#3, $\epsilon$ was set at higher levels of $\{0.3, 1, 8\}$ and $\delta$ was kept the same. NIST PSCR increased the $\epsilon$ values due to the increased number of variables in the PUMS data used for Match \#3. Additionally, NIST PSCR asked contestants to generate multiple differentially private data sets for each match. Matches \#1 and \#2 required three synthetic replicates while Match \#3 required five synthetic replicates. All contestants divided $\epsilon$ equally across each data set, which, per Theorem \ref{thm:comp}, Composition, resulted in using $\epsilon / m$ for each single data set. Accordingly, the $\epsilon$ used \textit{per} data set in Matches \#1 and \#2 was $\{0.00\overline{3}, 0.0\overline{3}, 0.\overline{3}\}$ and for Match \#3 the per data set $\epsilon$ was $\{0.06, 0.2, 1.6\}$.

\begin{table}[!htb]
    \def\arraystretch{1.2}
    \caption{NIST PSCR Differential Privacy Synthetic Data Challenge Results from the three Marathon Matches.}
    \label{tab:NIST-results}
    \centering\small
    \begin{tabular}{ C{1.2cm} | C{4.20cm} | C{4.20cm} | C{4.20cm}}
    \hline
        \textbf{Rank} & \textbf{Match \#1 }  & \textbf{Match \#2} & \textbf{Match \#3} \\
    \hline
          1 & \makecell{Team pfr \\ (Sec. \ref{subsub:pfr})} & \makecell{Team pfr \\ (Sec. \ref{subsub:pfr})} & \makecell{Team RMcKenna \\ (Sec. \ref{subsub:rmckenna})} \\
    \hline
          2 & \makecell{Team DPSyn \\ (Sec. \ref{subsub:dpsyn})} & \makecell{Team DPSyn \\ (Sec. \ref{subsub:dpsyn})} & \makecell{Team DPSyn \\ (Sec. \ref{subsub:dpsyn})} \\
    \hline
          3 & \makecell{Team RMcKenna \\ (Sec. \ref{subsub:rmckenna})} & \makecell{Team PrivBayes \\ (Sec. \ref{subsub:privbayes})} & \makecell{Team PrivBayes \\ (Sec. \ref{subsub:privbayes})} \\
    \hline
          4 & \makecell{Team UCLANESL \\ (Sec. \ref{subsub:ucla})} & \makecell{Team RMcKenna \\ (Sec. \ref{subsub:rmckenna})} & \makecell{Team Gardn999 \\ (Sec. \ref{subsub:gardner})} \\
    \hline
          5 & \makecell{Team PrivBayes \\ (Sec. \ref{subsub:privbayes})}  & \makecell{Team Gardn999 \\ (Sec. \ref{subsub:gardner})} & \makecell{Team UCLANESL \\ (Sec. \ref{subsub:ucla})}  \\          
    \hline
    \end{tabular}
\end{table}

\begin{figure}[!htb]
    \centerline{\includegraphics[width=6in]{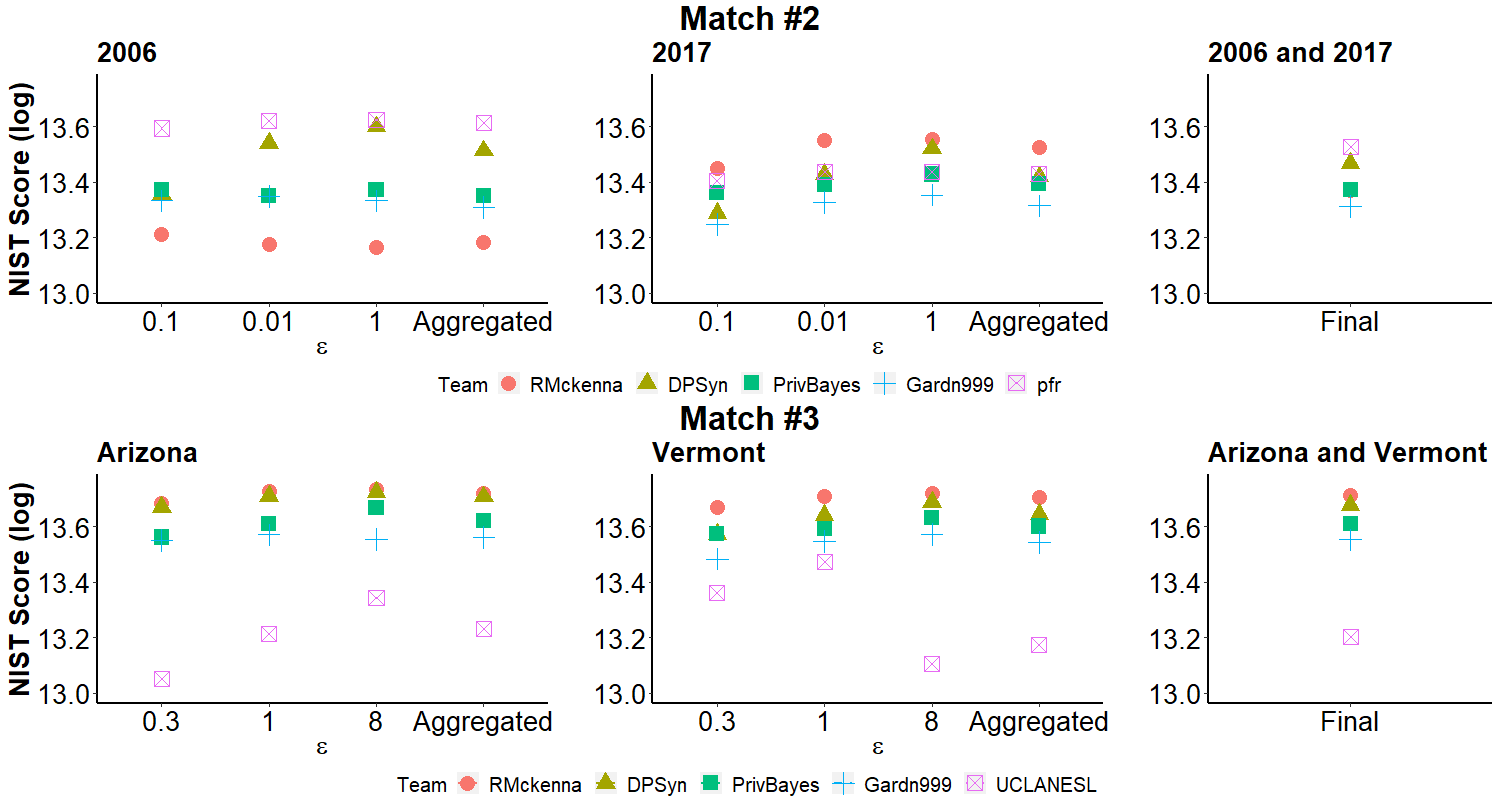}}
    \caption{The average NIST PSCR Differential Privacy Synthetic Data Challenge score results on a log scale for Matches \#2 (2006 and 2017) and \#3 (Arizona and Vermont).}\label{fig:NIST_score}
\end{figure}

Table \ref{tab:NIST-results} lists the team ranks while Figure \ref{fig:NIST_score} shows the numerical results of the NIST scoring metrics for the challenge matches. There are a total of six teams that ranked in the Top 5 throughout the three matches. Note that teams Gardn999, UCLANESL, and pfr did not compete in Matches \#1, \#2, and \#3, respectively, so their absence from the Top 5 for each match was not due to scoring lower than fifth. In Figure \ref{fig:NIST_score}, we plotted the original NIST PSCR specific utility score values, where larger values indicate better utility. (Note that in the following utility evaluation, we rescale the NIST scores, such that a smaller value indicates high utility.)

We focus our comparison on Matches \#2 and \#3, since Match \#1 used one of the two metrics and the same underlying data as Match \#2. Examining the results, Match \#2 shows fairly flat scores whereas Match \#3 scores slightly increased with larger $\epsilon$ values (except for Team UCLANESL on the Vermont data). The lack of any trend in Match \#2 for increasing $\epsilon$ is mostly likely due to the small $\epsilon$ values used for scoring. It may also be due to the fact that the data used in Match $\#2$ had more structural zeros. For future differentially private data challenges, using a wider range of $\epsilon$ would help verify empirically if the methods demonstrate a trend towards higher accuracy as $\epsilon$ grows. Algorithms that do not eventually asymptote towards maximal accuracy as privacy-loss goes to infinity suggest inherent flaws in the mechanism and should be utilized cautiously. A data practitioner seeking to determine suitable algorithms needs to test potential methods at a wide enough range of privacy-loss budget values to capture the risk-utility trade-off curve.

\subsection{Evaluation of Algorithms Using Quality Metrics}\label{sec:utilty_eval}
\begin{table}[!htb]
    \def\arraystretch{1.2}
    \caption{Match \#2 marginal distribution utility results for the top 5 scoring competitors. Results are averaged across multiple synthetic data sets for each test data set (2006 and 2017). Best results for each measure are in bold.}\label{tab:match2_marg}
    \centering\small
    \begin{tabular}{ C{1cm} | C{2.5cm} | C{3.5cm} | C{3.25cm} | C{3.25cm}}
  \hline
  $\epsilon$ & Entry & NIST PSCR `Classification' & $\chi^2_{pval}$ Mean & $KS_{pval}$ Mean \\ \hline
  0.01 & DPSyn & 0.35 & $<$0.01 & 0.00 \\ 
  0.01 & Gardn999 & 0.35 & 0.03 & 0.00 \\ 
  0.01 & pfr & \textbf{0.28} & $<$0.01 & 0.00 \\ 
  0.01 & PrivBayes & 0.32 & \textbf{0.69} & \textbf{0.33} \\ 
  0.01 & RMcKenna & 0.36 & $<$0.01 & 0.00 \\ 
  \hline
  0.10 & DPSyn & \textbf{0.27} & $<$0.01 & 0.00 \\ 
  0.10 & Gardn999 & 0.33 & 0.07 & 0.00 \\ 
  0.10 & pfr & 0.28 & 0.02 & $<$0.01 \\ 
  0.10 & PrivBayes & 0.31 & \textbf{0.69} & \textbf{0.33} \\ 
  0.10 & RMcKenna & 0.36 & $<$0.01 & 0.00 \\ 
  \hline
  1.00 & DPSyn & \textbf{0.25} & $<$0.01 & 0.00 \\ 
  1.00 & Gardn999 & 0.33 & \textbf{0.28} & 0.03 \\ 
  1.00 & pfr & 0.28 & 0.22 & \textbf{0.33} \\ 
  1.00 & PrivBayes & 0.29 & 0.23 & 0.08 \\ 
  1.00 & RMcKenna & 0.36 & $<$0.01 & 0.00 \\
  \hline
\end{tabular}
\end{table}

We now break down each algorithm's performance based on our three categories. Table \ref{tab:match2_marg} gives the results from Match \#2 for the marginal distribution metrics. We adjusted the original values for the NIST PSCR ``classification" metric to range from 0 to 1 with optimal scores equal to 0. The p-value metrics also range from 0 to 1, but 1 is the optimal value. We see that pfr and DPSyn perform well on the NIST PSCR task, while PrivBayes performs significantly better on the univariate comparisons until $\epsilon = 1$. Given the nature of the data, which includes many categorical variables with hundreds of levels and fine-grained \textit{Date} variables, we are not surprised to see poor performance of most algorithms on the univariate measures. Perhaps more surprisingly, PrivBayes performs well on those measures, but does not claim the top spot when comparing 3-way marginals. This suggests finer levels of algorithmic tweaking occurred for the others, pfr and DPSyn in particular, such that they matched 3-way marginals without matching 1-way marginals. For example, DPSyn post-processed the data to ensure 3-way marginal consistency, and pfr exclusively targeted accuracy on 3-way marginals. Because \textit{PrivBayes} uses unsupervised pre-processing, it appears to have primarily learned the univariate distributions.

\begin{table}[!htb]
    \def\arraystretch{1.2}
    \caption{Match \#2 joint distribution utility results for the top 5 scoring competitors. Results are averaged across multiple synthetic data sets for each test data set (2006 and 2017). Best results for each measure are in bold.}\label{tab:match2_joint}
    \centering\small
    \begin{tabular}{ C{0.75cm} | C{2cm} | C{2cm} | C{2cm} | C{2cm} | C{2cm} | C{2cm}}
  \hline
  $\epsilon$ & Entry & NIST PSCR `Clustering' & $pMSE_{ratio}$ (log) & $KS_{D}$ & $pMSE_{ratio}$ (log) & $KS_{D}$ \\ 
  \hline
  & & & \multicolumn{2}{c |}{CART} & \multicolumn{2}{c }{GLM} \\
  \hline
  0.01 & DPSyn & 0.43 & 6.50 & 0.99 & 10.35 & 0.94 \\ 
  0.01 & Gardn999 & 0.53 & 6.41 & 0.99 & 10.37 & 0.75 \\ 
  0.01 & pfr & \textbf{0.25} & \textbf{6.31} & \textbf{0.97} & 11.53 & 1.00 \\ 
  0.01 & PrivBayes & 0.42 & 6.51 & 0.99 & \textbf{7.56} & \textbf{0.53} \\ 
  0.01 & RMcKenna & 0.40 & 6.51 & 1.00 & 10.31 & 0.92 \\ 
  \hline
  0.10 & DPSyn & 0.29 & 6.48 & \textbf{0.98} & 10.26 & 0.89 \\ 
  0.10 & Gardn999 & 0.44 & 6.42 & 0.99 & 9.95 & 0.57 \\ 
  0.10 & pfr & \textbf{0.22} & \textbf{6.30} & 1.00 & 11.69 & 1.00 \\ 
  0.10 & PrivBayes & 0.40 & 6.48 & 0.99 & \textbf{7.57} & \textbf{0.52} \\ 
  0.10 & RMcKenna & 0.35 & 6.50 & 1.00 & 10.31 & 0.93 \\ 
  \hline
  1.00 & DPSyn & \textbf{0.21} & 6.43 & \textbf{0.97} & 10.22 & 0.87 \\ 
  1.00 & Gardn999 & 0.42 & 6.43 & 0.99 & 9.15 & 0.62 \\ 
  1.00 & pfr & 0.21 & \textbf{6.28} & 1.00 & 11.75 & 1.00 \\ 
  1.00 & PrivBayes & 0.39 & 6.48 & 0.99 & \textbf{8.46} & \textbf{0.54} \\ 
  1.00 & RMcKenna & 0.35 & 6.50 & 1.00 & 10.32 & 0.93 \\ 
  \hline
\end{tabular}
\end{table}

Next, we consider utility metrics measured on a larger number of variables jointly. The results are shown in Table \ref{tab:match2_joint}. When applying the discriminant-based utility metric algorithms from Section \ref{subsec:metric}, we implemented both CART and logistic regression with all main effects of the variables for estimating the predicted probabilities. We used the \verb;R; package \verb;rpart; for CART with a CP chosen by cross-validation. Since there were multiple synthetic data sets, we calculated the \textit{pMSE}-ratio and KS distance for each data set (using the same CP values across all data sets) and then averaged the results. For the \textit{pMSE}-ratio with the CART models, we generated 100 permutations to estimate the null mean \textit{pMSE}. Finally, we use a natural logarithmic transformation for the \textit{pMSE}-ratio values, since the values can rapidly increase on the tail. This means the optimal value for all metrics in Table \ref{tab:match2_joint} is 0. We bounded the NIST PSCR metric and the KS metrics between 0 and 1, and the \textit{pMSE}-ratio values are unbounded.

We see somewhat similar results in Table \ref{tab:match2_joint} as we saw in Table \ref{tab:match2_marg}. Entries by Team pfr consistently performed strongest on the NIST PSCR ``clustering" metric and the CART-based metrics. Conversely, the GLM modeled propensity score approaches assigned the highest value to Team PrivBayes. As expected, the GLM utility metrics are primarily driven by univariate distributional differences, which assign similar value as the p-value metrics. We also note the difference between Team DPSyn results using the NIST PSCR metric versus the \textit{pMSE}-ratio and SPECKS using the CART model. DPSyn clearly performs second best overall using the NIST PSCR metric, but its performance is average based on the \textit{pMSE}-ratio and SPECKS. Recall that the ``clustering" metric relied on randomly chosen subsets of one-third of the variables in the data and random subsets of those variables ranges. These results indicate that Team DPSyn better preserved lower order interactions than most other algorithms, but it did not noticeably better preserve aspects of the full joint distribution. In general, we see more differentiation on the lower order utility metrics, while the full joint metrics, based on the CART models, gave fairly similar scores for all algorithms. This reflects the complex nature of the original data, which included categorical variables with higher number of values, locations, and timestamps, and it suggests none of the competitors differentiated themselves on capturing the whole joint distribution.

\begin{table}[!htb]
    \def\arraystretch{1.2}
    \caption{Match \#3 marginal distribution utility results for the top 5 scoring competitors. Results are averaged across multiple synthetic data sets for each test data set (Arizona and Vermont). Best results for each measure are in bold.}\label{tab:match3_marg}
    \centering\small
    \begin{tabular}{ C{1cm} | C{2.5cm} | C{3.5cm} | C{3.25cm} | C{3.25cm}}
  \hline
  $\epsilon$ & Entry & NIST PSCR `Classification' & $\chi^2_{pval}$ Mean & $KS_{pval}$ Mean \\ \hline
  0.30 & DPSyn & 0.31 & 0.02 & 0.00 \\ 
  0.30 & Gardn999 & 0.32 & 0.02 & 0.02 \\ 
  0.30 & PrivBayes & 0.35 & 0.01 & 0.01 \\ 
  0.30 & RMcKenna & \textbf{0.17} & \textbf{0.09} & \textbf{0.06} \\ 
  0.30 & UCLANESL & 0.72 & 0.00 & 0.00 \\ 
  \hline
  1.00 & DPSyn & 0.23 & 0.05 & 0.12 \\ 
  1.00 & Gardn999 & 0.28 & 0.04 & 0.15 \\ 
  1.00 & PrivBayes & 0.33 & 0.02 & 0.03 \\ 
  1.00 & RMcKenna & \textbf{0.15} & \textbf{0.18} & \textbf{0.29} \\ 
  1.00 & UCLANESL & 0.53 & 0.00 & 0.00 \\ 
  \hline
  8.00 & DPSyn & 0.20 & 0.27 & 0.55 \\ 
  8.00 & Gardn999 & 0.26 & 0.16 & 0.55 \\ 
  8.00 & PrivBayes & 0.31 & 0.03 & 0.11 \\ 
  8.00 & RMcKenna & \textbf{0.14} & \textbf{0.29} & \textbf{0.73} \\ 
  8.00 & UCLANESL & 0.41 & 0.00 & 0.00 \\ 
  \hline
\end{tabular}
\end{table}

Similar to the NIST PSCR scores from Figure \ref{fig:NIST_score}, the joint utility metrics estimated from both classifiers for Match \#2 provide relatively flat values, likely due to the very small and narrow range of $\epsilon$ values. In other words, we do not see large increases in utility, or even increases at all for some algorithms, as $\epsilon$ increases. 
%As we will show with Match \#3, having a wider range of $\epsilon$ values and evaluating additional utility metrics provides additional levels of differentiation. 
In Section \ref{sec:change}, we also further discuss the flatness of the change in utility as the privacy budget increase.

Moving to Match \#3, Table \ref{tab:match3_marg} displays the results for the marginal distribution metrics for the top 5 contestants. The best performing algorithm from Matches \#1 and \#2, Team pfr, did not compete in Match \#3, so, unfortunately, we cannot compare its performance using this data. According to the first set of measures, Team RMcKenna outperformed the others by a significant margin, demonstrating that either their algorithm improved between matches or performed better on Match \#3 data than Match \#2 data. In contrast to Match \#2, we see consistency between the NIST PSCR 3-way marginal metric and the two univariate measures, which suggests that 1-way and 3-way marginals are more closely related on this data set or that competitors targeted both equally. The algorithms may have targeted the 3-way metric less, since NIST PSCR introduced a third scoring metric for this match.

\begin{table}[!htb]
    \def\arraystretch{1.2}
    \caption{Match \#3 joint distribution utility results for the top 5 scoring competitors. Results are averaged across multiple synthetic data sets for each test data set (Arizona and Vermont). Best results for each measure are in bold.}\label{tab:match3_joint}
    \centering\small
    \begin{tabular}{ C{0.75cm} | C{2cm} | C{2cm} | C{2cm} | C{2cm} | C{2cm} | C{2cm}}
  \hline
  $\epsilon$ & Entry & NIST PSCR `Clustering' & $pMSE_{ratio}$ (log) & $KS_{D}$ & $pMSE_{ratio}$ (log) & $KS_{D}$ \\ 
  \hline
  & & & \multicolumn{2}{c |}{CART} & \multicolumn{2}{c }{GLM} \\
  \hline
  0.30 & DPSyn & 0.15 & 6.48 & 0.97 & 9.37 & 0.80 \\ 
  0.30 & Gardn999 & 0.21 & 6.67 & 1.00 & 10.17 & 0.91 \\ 
  0.30 & PrivBayes & 0.23 & 6.01 & 0.99 & 6.30 & 0.33 \\ 
  0.30 & RMcKenna & \textbf{0.12} & \textbf{5.50} & \textbf{0.80} & \textbf{6.15} & \textbf{0.20} \\ 
  0.30 & UCLANESL & 0.57 & 6.81 & 1.00 & 11.17 & 1.00 \\ 
  \hline
  1.00 & DPSyn & 0.11 & 6.39 & 0.90 & 8.33 & 0.71 \\ 
  1.00 & Gardn999 & 0.18 & 6.66 & 0.99 & 9.25 & 0.68 \\ 
  1.00 & PrivBayes & 0.21 & 6.04 & 1.00 & 5.75 & 0.28 \\ 
  1.00 & RMcKenna & \textbf{0.09} & \textbf{4.46} & \textbf{0.56} & \textbf{4.57} & \textbf{0.23} \\ 
  1.00 & UCLANESL & 0.42 & 6.81 & 1.00 & 10.98 & 0.98 \\
  \hline
  8.00 & DPSyn & 0.09 & 6.30 & 0.81 & 7.54 & 0.71 \\ 
  8.00 & Gardn999 & 0.17 & 6.64 & 0.98 & 5.83 & 0.28 \\ 
  8.00 & PrivBayes & 0.18 & 6.00 & 0.99 & 5.63 & \textbf{0.25} \\ 
  8.00 & RMcKenna & \textbf{0.07} & \textbf{4.97} & \textbf{0.59} & \textbf{2.23} & 0.26 \\ 
  8.00 & UCLANESL & 0.35 & 6.80 & 0.99 & 10.80 & 0.91 \\
  \hline
\end{tabular}
\end{table}

\begin{table}[!htb]
    \def\arraystretch{1.2}
    \caption{Match \#3 Correlation utility results for the top 5 scoring competitors. Results are averaged across multiple synthetic data sets for each test data set (Arizona and Vermont). Best results for each measure in bold.}\label{tab:match3_corr}
    \centering\small
    \begin{tabular}{ C{0.75cm} | C{2cm} | C{2cm} | C{2cm} | C{2cm} | C{2cm} | C{2cm}}
  \hline
  $\epsilon$ & Entry & NIST PSCR `Regression' & CI Overlap & Std. $\hat \beta$ Diff. & CI Overlap & $\hat \beta$ Diff. \\ 
  \hline
  & & & \multicolumn{2}{c |}{Model 1} & \multicolumn{2}{c }{Model 2} \\
  \hline
  0.30 & DPSyn & 0.07 & \textbf{0.61} & 4.18 & 0.64 & 2.44 \\ 
  0.30 & Gardn999 & 0.25 & 0.61 & 3.75 & 0.50 & 2.26 \\ 
  0.30 & PrivBayes & 0.26 & 0.59 & 4.05 & \textbf{0.68} & \textbf{2.26} \\ 
  0.30 & RMcKenna & \textbf{0.10} & 0.55 & \textbf{2.71} & 0.58 & 2.46 \\ 
  0.30 & UCLANESL & 0.22 & - & - & - & - \\ 
  \hline
  1.00 & DPSyn & 0.05 & 0.39 & 4.78 & 0.63 & \textbf{2.70} \\ 
  1.00 & Gardn999 & 0.22 & 0.52 & \textbf{2.89} & \textbf{0.63} & 3.35 \\ 
  1.00 & PrivBayes & 0.27 & \textbf{0.63} & 5.57 & 0.56 & 3.19 \\ 
  1.00 & RMcKenna & \textbf{0.04} & 0.30 & 13.05 & 0.53 & 2.77 \\ 
  1.00 & UCLANESL & 0.28 & 0.52 & 9.70 & - & - \\ 
  \hline
  8.00 & DPSyn & \textbf{0.02} & 0.77 & 2.51 & 0.66 & 3.70 \\ 
  8.00 & Gardn999 & 0.24 & \textbf{0.83} & \textbf{2.04} & \textbf{0.71} & \textbf{1.91} \\ 
  8.00 & PrivBayes & 0.26 & 0.61 & 4.85 & 0.70 & 2.71 \\ 
  8.00 & RMcKenna & 0.04 & 0.67 & 2.81 & 0.58 & 4.03 \\ 
  8.00 & UCLANESL & 0.25 & 0.52 & 15.56 & - & - \\
  \hline
\end{tabular}
\end{table}

Table \ref{tab:match3_joint} provides the results for each algorithm using the joint distributional metrics. In this match, we see general agreement across the joint utility metrics and between the joint and marginal metrics. Team RMcKenna performs the best on almost every metric and for every level of $\epsilon$. Overall, the contestants (the four teams who participated in both Match \#2 and \#3) scored much higher on all the marginal and joint metrics for Match \#3, suggesting improvements of algorithms, an easier data set to synthesize, or some combination of the two.

Finally, we summarize the correlation utility results in Table \ref{tab:match3_corr}. These results include the NIST PSCR metric based on ranking cities by gender wage gap and the mean confidence interval overlap and standardized coefficient differences for all coefficients from two regression models (one logistic, one Poisson). Utility values could not be calculated for Team UCLANESL for certain combinations of $\epsilon$ and regression models, because it produced synthetic data sets with no variation in our pre-selected outcome variables. The correlation utility results differ from the marginal and joint measures, and there is no consistent best performing algorithm across the metrics. Team RMcKenna performs the best on the NIST PSCR score, apart from $\epsilon = 8$ whereas teams Gardn999 and PrivBayes generally rank the best on the regression metrics. The varied performance on the correlation metrics reflects the highly specific nature of these metrics, and the fact that the regressions were not part of the workload. Because the NIST PSCR ``regression" metric was public, we can clearly see that DPSyn and RMcKenna allocated more privacy budget towards preserving the correlations of gender and wage within cities, yet neither of these performed the best on the regression models.

One may contend that it is unfair to evaluate algorithms based on highly specific utility metrics that were not part of the workload, but this leads to an observation that more general pre-processing methods contribute to better results on unexpected specific tasks. In some cases, data providers may not know all of the planned uses for synthetic data. These results suggests that if we do not know what models a data user plans to estimate, we are better off using a general pre-processing step that tries to preserve all high correlations. We clearly see differences in how the teams used pre-processing to target certain workloads. Teams RMcKenna and DPSyn prioritized certain 1-, 2-, or 3-way marginals while PrivBayes did not because it relies on unsupervised pre-processing. Conversely, Gardn999 applied only general pre-processing to identify correlations across the whole data.

UCLANESL seem to have prioritized the NIST PSCR ``regression" scoring, because it performed much better on that metric than the other two NIST PSCR measures. Unfortunately, their approach did not translate to other specific models, and in some cases did not even produce data capable of estimating the models. Apart from UCLANESL, it is important to note that the other four entries performed fairly well on preserving correlations in the data. On average, the coefficients in the regression models had above 60\% CI overlap with the original data, and, in some cases, 70\% or even 80\%. These results indicate that differentially private synthetic data, particularly at higher levels of $\epsilon$, can perform reasonably well at preserving tasks that a statistician would typically conduct on this type of data.

\subsection{Change in Quality as Epsilon Increases}\label{sec:change}
Ideally, differentially private algorithms should improve the quality of their output as $\epsilon$ increases, since a higher privacy loss implies less noise has been added to the data. However, synthetic data with substantial pre- or post-processing may obscure the $\epsilon$-quality trade-off. While we would prefer an algorithm that performs better at any value of $\epsilon$ over an algorithm that performs worse, in general, we prefer algorithms that exhibit the natural utility-privacy loss trade-off curve over those that do not change accuracy with $\epsilon$.

\begin{figure}[!htb]
    \centerline{\includegraphics[width=6.5in]{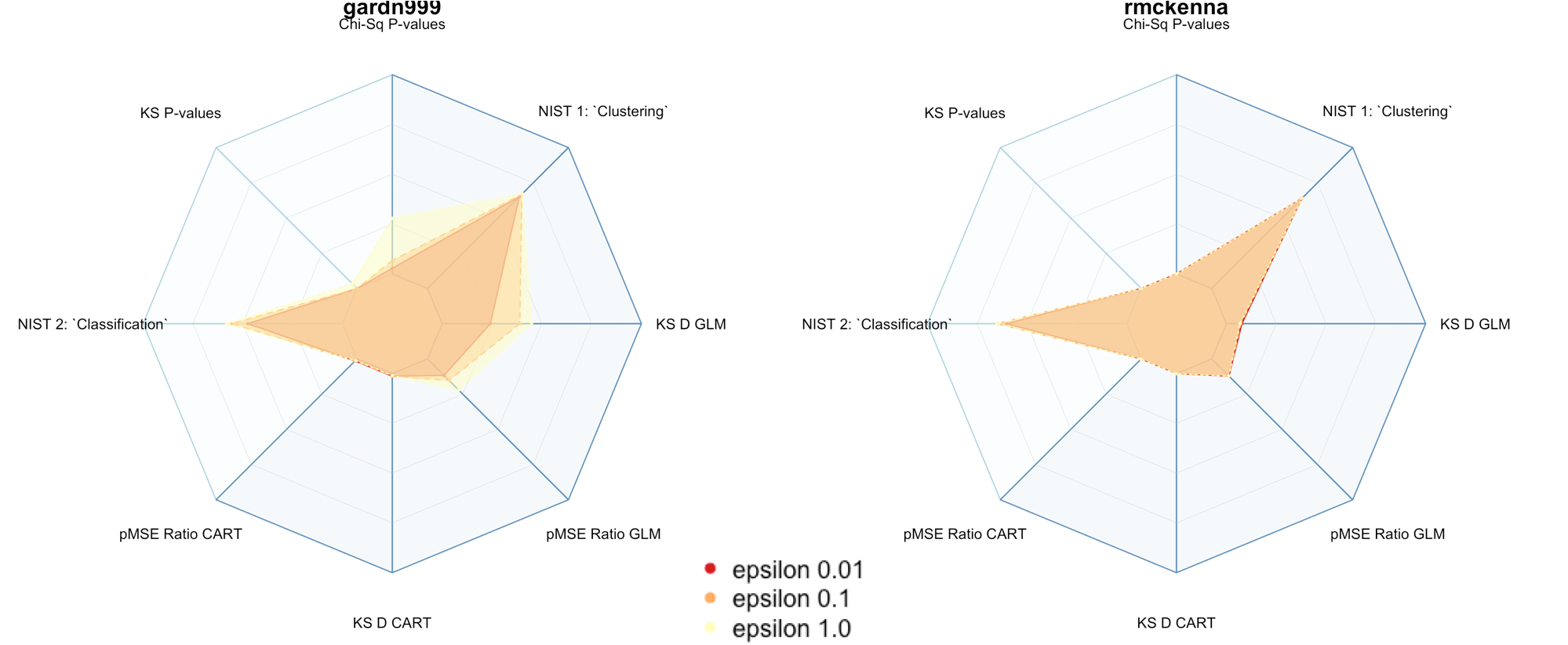}}
    \caption{Utility plot for Team RMcKenna's and Team Gardn999's results in Match \#2.}\label{fig:radar_match2}
\end{figure}

\begin{figure}[!htb]
    \centerline{\includegraphics[width=6.5in]{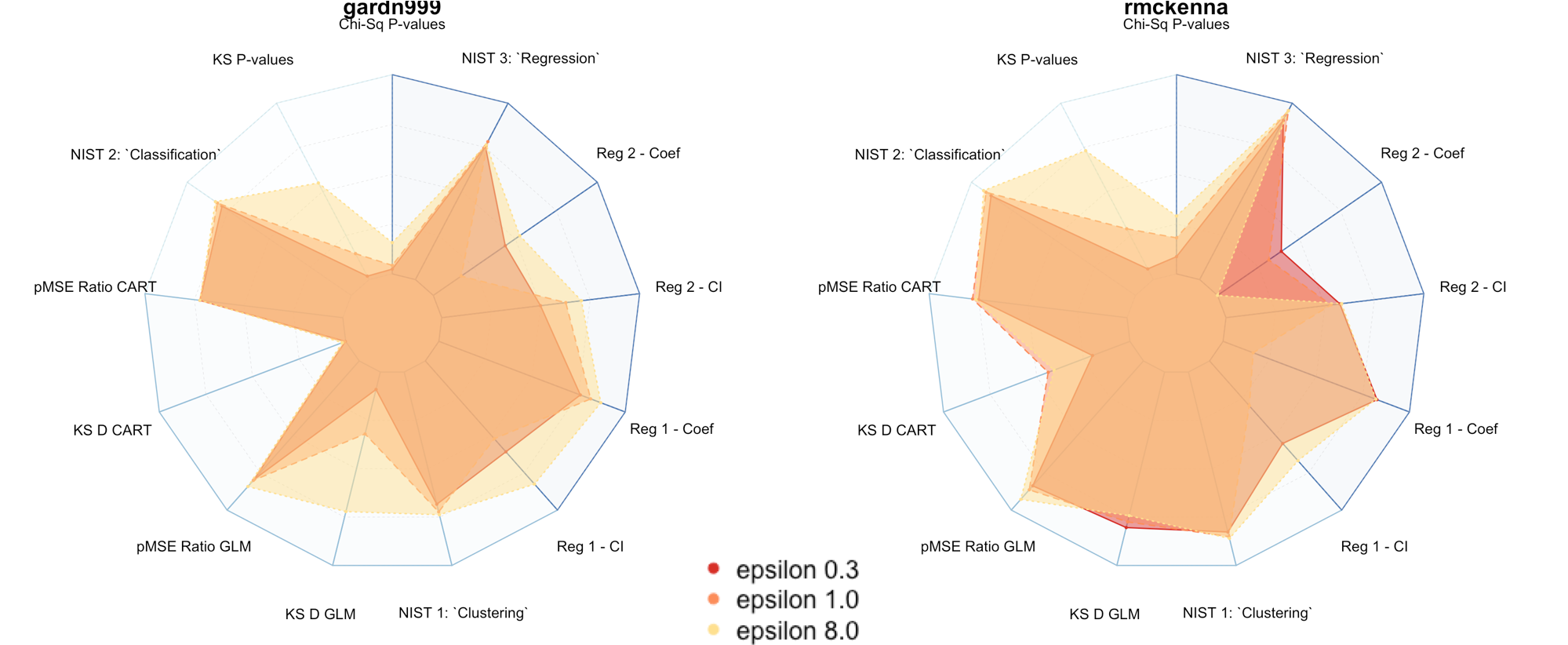}}
    \caption{Utility plot for Team RMcKenna's and Team Gardn999's results in Match \#3.}\label{fig:radar_match3}
\end{figure}

In order to visualize the change in accuracy as $\epsilon$ increases, we adapt the radarchart utility plots used by \cite{arnold2020really}. This graphic allows us to visualize all the utility metrics simultaneously and perform a relative comparison to see which algorithms performed well on what types of metrics. Figures \ref{fig:radar_match2} and \ref{fig:radar_match3} show the utility plot for two competitors', teams RMcKenna and Gardn999, and their results on Matches \#2 and \#3. Each plot displays the utility for a given algorithm on each metric, and the various shaded blue segments of the plot represent the marginal, joint, and correlation utility categories from earlier. The different orange shaded areas correspond to each level of $\epsilon$, with darker values indicating lower $\epsilon$. A larger area coverage indicates increasing utility values for the corresponding metrics. These charts offer a nice and easy way to quickly visualize and synthesize a lot of information both in terms of the different metrics and the different levels of $\epsilon$.

Figure \ref{fig:radar_match2} displays that little change occurred in the scores for the three different values of $\epsilon$ in Match \#2. Team RMcKenna performed almost identically for all levels of $\epsilon$ while Team Gardn999 exhibited slight improvement as $\epsilon$ increased. This aligns with what we noted before, that the results at different noise levels were mostly indiscernible, likely due to the low $\epsilon$ values in this match. We see that while teams RMcKenna and Gardn999 performed similarly on the NIST PCSR metrics, Team Gardn999 performed better on some of the other metrics.

By contrast, Figure \ref{fig:radar_match3} shows that, for most of the metrics in Match \#3, the utility increases with higher levels of $\epsilon$. The only exception is Team RMcKenna's results for the standardized coefficient differences for regression model 2 (as well as very slight exceptions for the KS D metrics). But, with a utility metric that measures accuracy on a highly specific value such as a regression coefficient, it is understandable that a given instance might have more variability in its performance. While the noise on average is smaller for higher $\epsilon$, the noise still comes from random distributions that vary from instance to instance. In particular, the regression utility metrics appear sensitive to a few bad replicates. For example, Team RMcKenna's utility at $\epsilon = 1$ appears to be affected by some poor synthetic replicates. Most of Team Gardn999's results display steady improvement as $\epsilon$ increases. We also see from these charts that Team RMcKenna achieved stronger utility on the marginal and joint metrics, but Team Gardn999 performed better overall on the correlation metrics.

Without showing the rest of the figures, we find in general that the non-parametric algorithms (teams DPSyn, pfr, and Gardn999) were more likely to show steady improvement as $\epsilon$ increased whereas the parametric synthesis algorithms (teams RMcKenna, PrivBayes, and UCLANESL) often showed either no change or sometimes decrease in utility. Overall, Team Gardn999 demonstrated the most direct relationship between $\epsilon$ and utility. Two reasons likely explain these findings. First, the parametric algorithms add noise in a less direct fashion, drawing values from a high-dimensional distribution that must first be approximated. The non-parametric algorithms, on the other hand, add noise directly to the marginals. This suggests that while parametric models can produce good results, such as Team RMcKenna, they cannot be as easily changed using different values of $\epsilon$. Second, methods which involve extensive post-processing introduce additional noise into the data, such that the risk-utility trade-off does not depend directly on $\epsilon$. Team Gardn999 used the least post-processing among all competitors, and we see that their algorithm produced the most direct relationship between privacy loss and utility.

\section{Conclusions and Future Recommendations}\label{sec:discussion}
In this paper, we reviewed and evaluated the top methods from PSCR’s NIST Differential Privacy Synthetic Data Challenge methods on a wide range of utility metrics. Our evaluation is the first comparative work, to the best of our knowledge, that assesses a variety of DP synthetic data generation mechanisms applied to complex real-world data sets and gives recommendations concerning their accuracy and ease of implementation. For simplicity, we summarize the practical findings of this paper in the following:
\begin{enumerate}
    \item The best performing differentially private synthetic data algorithms used pre-processing and budget allocation based on public data and subject matter utility criteria. This significantly reduced the output domain of the synthetic data, allowing the use of basic mechanisms applied to a simplified histogram.
    \item Data practitioners can apply pre-processing and budget allocation generally, such as capturing all highly correlated variables in the data, or narrowly, such as preserving specific relationships as we saw with the Match \#3 ``regression" tasks. These choices present a general-specific utility trade-off.
    \item Non-parametric and parametric algorithms offer an implementation trade-off between requiring extensive pre-processing from using public data versus requiring significant computational capabilities. Data practitioners may choose one or the other based on their available resources.
    \item Parametric methods often do not improve as $\epsilon$ increases because post-processing creates error that affects the privacy-utility trade-off and does not diminish as $\epsilon$ increases. Data providers, who desire a straightforward risk-utility trade-off curve, should consider non-parametric methods with minimal post-processing.
    \item GANs achieved much lower utility than the simpler methods. These methods also require significant computational burdens, which typical data practitioners might not possess. Apart from the computational issues, we believe UCLANESL did not perform well due to the extensive clipping and the termination of the algorithm when the privacy budget was expended. The GAN process likely did not optimize by the time it was aborted.
    \item Match \#3 showed that at higher levels of $\epsilon$, some algorithms produced fairly high quality and usable synthetic data. This indicates that differentially private synthetic data shows promise as a SDC approach.
\end{enumerate}
We recommend a wider range of the privacy-loss budget to be explored for future DP data competitions. As seen in Match \#2, we saw a lack of an asymptotic trend as $\epsilon$ increased, which made it difficult to learn from this match. The challenge also privileged competitors' who honed their algorithms towards specific metrics, such as 3-way marginals. A future challenge with more general scoring metrics might lead to competitors submitting more generalized and useful algorithms. Additionally, the NIST PSCR data sets used in the competition were very complex compared to what is typically seen in literature, and these results suggest such complex data require more privacy-loss budget for greater accuracy. Elements such as structural missing values in the Match \#2 data, e.g., some emergency calls do not involve an officer dispatch, or the large number of variables in Match \#3 greatly increased the difficulty of providing accurate differentially private synthetic data.

The various utility metric algorithms offered mixed evaluations on which differentially private data synthesis method performed best.
% the NIST PSCR Data Challenge outcome depended on the choice of classifier and the NIST PSCR evaluation standards used in different matches. 
These rankings point to the difference in what the utility metrics measure. Given the results, we suggest that data practitioners wishing to select DP algorithms for releasing data should use a suite of metrics for a more informative evaluation of the quality of the algorithms. 
%For instance, the discriminator approaches measure different types of accuracy depending on the classifier used, and should be selected based on the desired utility improvements from the synthetic data compared to the original data. 
%For future work, we would explore and investigate additional loss functions and classification methods (e.g., SVM) on discriminator approaches. Specifically, we would conduct a thorough investigation of what data qualities and features are being measured. However, there is a concern for using certain classifiers such as SVM that are computationally more expensive. Since most DP methods are computationally intense, the utility metrics should ideally be calculated using very little of the computational resources, which might eliminate SVM and other classifiers in practice. 
Besides the quality metrics we described in Section \ref{sec:DP-Metric}, there are other ways to extensively evaluate differentially private methods such as DPBench \citep{hay2016principled}, which could additionally be deployed.
% Future challenges could be developed to facilitate using the DPBench evaluation process. This would expand our current knowledge of how to better evaluate differentially private synthetic data.
This paper is the first to offer practical insights on the relative usefulness of different DP synthetic data algorithms using real data. Future work and applications should seek to continue comparing mechanisms using real data in order to inform data providers wishing to generate differentially private synthetic data.

\section*{Acknowledgments}
\noindent This research is supported by the National Institute of Standards and Technology Public Safety Communications Research (PSCR) Division under P19-774-0002 and \\
1333ND18DNB630011-1333ND19FNB775355. PSCR is the primary federal laboratory conducting research, development, testing and evaluation for public safety communications technologies.

The publication has been assigned the Los Alamos National Laboratory identifier LA-UR-19-31841.

We would like to thank Marcel Neunhoeffer and Christian Arnold for providing code and advice on the radar chart visualizations.
\bibliographystyle{apa}
\setlength{\bibhang}{0pt}
\bibliography{ref}

\end{document}